%% file: EPV_JASA.tex
\documentclass[12pt]{article}
\usepackage{subfiles, titling, setspace, authblk}
\usepackage{todonotes}
\usepackage{enumitem}
\usepackage{algpseudocode}
\usepackage{algorithm}
\usepackage{booktabs}
\usepackage{tikz}
\usepackage{graphicx}
\usepackage[caption=false]{subfig}
\usepackage{multirow}
\usepackage{array}

\usetikzlibrary{xyhoopsepv}

\usepackage{amsmath, amsfonts, mathrsfs, hyperref} 

\usepackage[font={small,it}]{caption}

\usepackage[default]{jasa_harvard}    
\usepackage{JASA_manu}
\newcommand{\citep}{\cite}
\newcommand{\citet}{\citeasnoun}

\DeclareMathAlphabet{\mathpzc}{OT1}{pzc}{m}{it}

	\newcommand{\E}{\mathbb{E}} 
	\newcommand{\prob}{\mathbb{P}}
	
    \newcommand{\Fz}{\mathcal{F}^{(Z)}}
    
	\newcommand{\R}{\mathbb{R}}
	\newcommand{\Z}{\mathcal{Z}}
	\newcommand{\N}{\mathcal{N}}
\newcommand{\Ss}{\mathbb{S}}
\newcommand{\bz}{\mathbf{z}}
\newcommand{\norm}{\mathcal{N}}
\newcommand{\Cset}{\mathcal{C}}

\newcommand{\bTh}{\boldsymbol{\Theta}}
\newcommand{\shot}{\textrm{s}}

	\newtheorem{theorem}{Theorem}[section]

	\newenvironment{definition}[1][Definition]{\begin{trivlist}
		\item[\hskip \labelsep {\bfseries #1}]}{\end{trivlist}}
	
	\newenvironment{remark}[1][Remark]{\begin{trivlist}
		\item[\hskip \labelsep {\bfseries #1}]}{\end{trivlist}}

	\newcommand{\qed}{\nobreak \ifvmode \relax \else
     	\ifdim\lastskip<1.5em \hskip-\lastskip
     	\hskip1.5em plus0em minus0.5em \fi \nobreak
     	\vrule height0.75em width0.5em depth0.25em\fi}

\title{A Multiresolution Stochastic Process Model for Predicting Basketball Possession Outcomes}

\author[1]{Daniel Cervone}
\author[2]{Alex D'Amour}
\author[3]{Luke Bornn}
\author[4]{Kirk Goldsberry}
\date{}

\affil[1]{Center for Data Science, New York University, New York, NY 10003} 
\affil[2]{Department of Statistics, Harvard University, Cambridge, MA 02138} 
\affil[3]{Department of Statistics and Actuarial Science, Simon Fraser University, Burnaby, BC, Canada} 
\affil[4]{Institute for Quantitative Social Science, Harvard University, Cambridge, MA 02138}

\begin{document} 

\pagenumbering{gobble}

\maketitle

\mbox{}
\vspace*{2in}
\begin{center}
\textbf{Author's Footnote:}
\end{center}
Daniel Cervone (email: \texttt{dcervone@nyu.edu}) is Moore-Sloan Data Science Fellow at New York University. Alex D'Amour (\texttt{damour@fas.har\-vard.edu}) is PhD candidate, Statistics Department, Harvard University. Luke Bornn (\texttt{lbornn@sfu.ca}) is Assistant Professor, Department of Statistics and Actuarial Science, Simon Fraser University. Kirk Goldsberry (\texttt{kgoldsberry@fas.harvard.edu}) is Visiting Scholar, Center for Geographic Analysis, Harvard University. Daniel Cervone is funded by a fellowship from the Alfred P. Sloan and Betty Moore foundations. Luke Bornn is funded by DARPA Award FA8750-14-2-0117, ARO Award W911NF-15-1-0172, an Amazon AWS Research Grant, and the Natural Sciences and Engineering Research Council of Canada.

The authors would like to thank Alex Franks, Andrew Miller, Carl Morris, Natesh Pillai, and Edoardo Airoldi for helpful comments, as well as STATS LLC in partnership with the NBA for providing the optical tracking data. The computations in this paper were run on the Odyssey cluster supported by the FAS Division of Science, Research Computing Group at Harvard University.

\begin{center}
\textbf{Abstract}
\end{center}
Basketball games evolve continuously in space and time as players
constantly interact with their teammates, the opposing team, and the
ball. However, current analyses of basketball outcomes rely on
discretized summaries of the game that reduce such interactions to
tallies of points, assists, and similar events. In this paper, we
propose a framework for using optical player tracking data to
estimate, in real time, the expected number of points obtained by the
end of a possession. This quantity, called \textit{expected possession
  value} (EPV), derives from a stochastic process model for the
evolution of a basketball possession. We model this process at
multiple levels of resolution, differentiating between continuous,
infinitesimal movements of players, and discrete events such as shot
attempts and turnovers. Transition kernels are estimated using
hierarchical spatiotemporal models that share information across
players while remaining computationally tractable on very large data
sets. In addition to estimating EPV, these models reveal novel
insights on players' decision-making tendencies as a function of their
spatial strategy. A data sample and R code for further exploration of our model/results are available in the repository \url{https://github.com/dcervone/EPVDemo}.

\vspace*{.3in}



\pagenumbering{arabic}
\section{Introduction}
	\label{sec:Intro}
	\subfile{EPV_intro.tex}



\section{Multiresolution Modeling}
	\label{sec:Multiresolution}
    \subfile{multires2.tex}
\section{Transition Model Specification}
	\label{sec:Macro}
	\subfile{EPV_macro.tex}

\section{Hierarchical Modeling and Inference}
	\label{sec:Computation}

\subfile{EPV_computation.tex}

\section{Results}
	\label{sec:Results}

\subfile{EPV_results.tex}

\section{Discussion}
	\label{sec:Discussion}
	\subfile{EPV_discuss.tex}

\newpage
\appendix

\makeatletter   
 \renewcommand{\@seccntformat}[1]{APPENDIX~{\csname the#1\endcsname}.\hspace*{1em}}
 \makeatother


\section{Additional Technical Details}
	\label{sec:fullspec}

\subfile{EPV_fullspec.tex}

\section{Data and Code}
	\label{sec:supplement}

The Git repository \url{https://github.com/dcervone/EPVDemo} contains a one game sample of optical tracking data (csv), along with R code for visualizing model results and reproducing EPV calculations. Pre-computed results of computationally-intensive steps are also included as Rdata files, and can be loaded to save time and resources. A reproducible knitr tutorial, \texttt{EPV\_demo.Rnw}, introduces the data and demonstrates core code functionality.

\singlespacing
\bibliographystyle{ECA_jasa}
\bibliography{Refs}


\end{document}

%% file: EPV_intro.tex
Basketball is a fast-paced sport, free-flowing in both space and time, in which players' actions and decisions continuously impact their teams' prospective game outcomes. Team owners, general managers, coaches, and fans all seek to quantify and evaluate players' contributions to their team's success. However, current statistical models for player evaluation such as ``Player Efficiency Rating'' \cite{hollinger2004pro} and ``Adjusted Plus/Minus'' \cite{omidiran2011pm} rely on highly reductive summary statistics of basketball games such as points scored, rebounds, steals, assists---the so-called ``box score'' summary of the game. Such models reflect the fact that up until very recently, data on basketball games were only available in this low level of resolution. Thus previous statistical analyses of basketball performance have overlooked many of the high-resolution motifs---events not measurable by such aggregate statistics---that characterize basketball strategy. For instance, traditional analyses cannot estimate the value of a clever move that fools the defender, or the regret of skipping an open shot in favor of passing to a heavily defended teammate. The advent of player tracking data in the NBA has provided an opportunity to fill this gap.

\subsection{Player-Tracking Data}
In 2013 the National Basketball Association (NBA), in partnership with data provider STATS LLC, installed optical tracking systems in the arenas of all $30$ teams in the league. The systems track the exact two-dimensional locations of every player on the court (as well as the three-dimensional location of the ball) at a resolution of 25Hz, yielding over 1 billion space-time observations over the course of a full season. 

Consider, for example, the following possession recorded using this player tracking system. This is
a specific Miami Heat possession against the Brooklyn Nets from the second quarter of a game on November 1, 2013, chosen arbitrarily among those during which LeBron James (widely considered the best NBA player as of 2014) handles the ball. 
\begin{figure}[h!]
\centering
\includegraphics[width=1.0\linewidth]{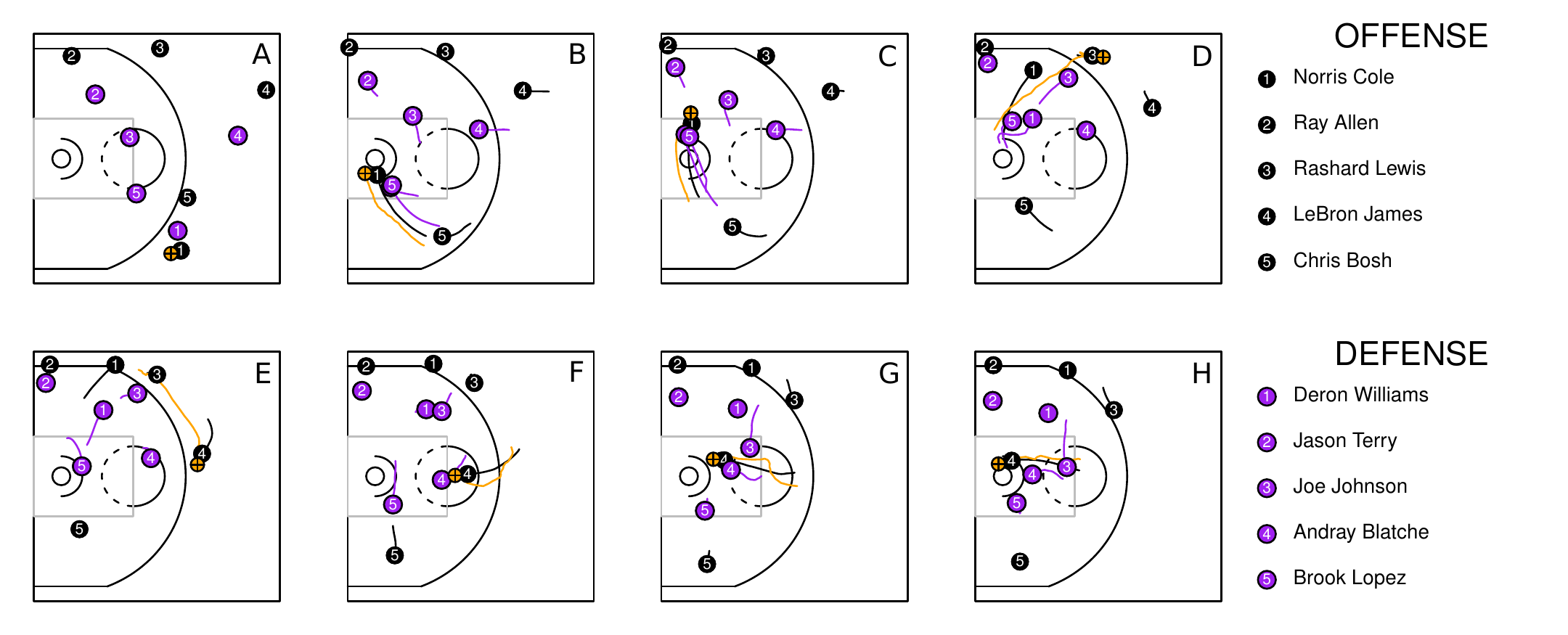}
\caption{Miami Heat possession against Brooklyn Nets. Norris Cole wanders into the perimeter (A) before driving toward the basket (B). Instead of taking the shot, he runs underneath the basket (C) and eventually passes to Rashard Lewis(D), who promptly passes to LeBron James (E). After entering the perimeter (F), James slips behind the defense (G) and scores an easy layup (H).}
\label{heat_poss}
\end{figure}
In this particular possession, diagrammed in Figure \ref{heat_poss}, point guard Norris Cole begins with possession of the ball crossing the halfcourt line (panel A). After waiting for his teammates to arrive in the offensive half of the court, Cole wanders gradually into the perimeter (inside the three point line), before attacking the basket through the left post. He draws two defenders, and while he appears to beat them to the basket (B), instead of attempting a layup he runs underneath the basket through to the right post (C). He is still being double teamed and at this point passes to Rashard Lewis (D), who is standing in the right wing three position. As defender Joe Johnson closes, Lewis passes to LeBron James, who is standing about 6 feet beyond the three point line and drawing the attention of Andray Blatche (E). James wanders slowly into the perimeter (F), until just behind the free throw line, at which point he breaks towards the basket. His rapid acceleration (G) splits the defense and gains him a clear lane to the basket. He successfully finishes with a layup (H), providing the Heat two points.
%

\subsection{Expected Possession Value}
Such detailed data hold both limitless analytical potential for basketball spectators and new methodological challenges to statisticians. Of the dizzying array of questions that could be asked of such data, we choose to focus this paper on one particularly compelling quantity of interest, which we call \textit{expected possession value} (EPV), defined as
the expected number of points the offense will score on a particular possession conditional on that possession's evolution up to time $t$.

\begin{figure}[h!]
\centering
\includegraphics[width=1.0\linewidth]{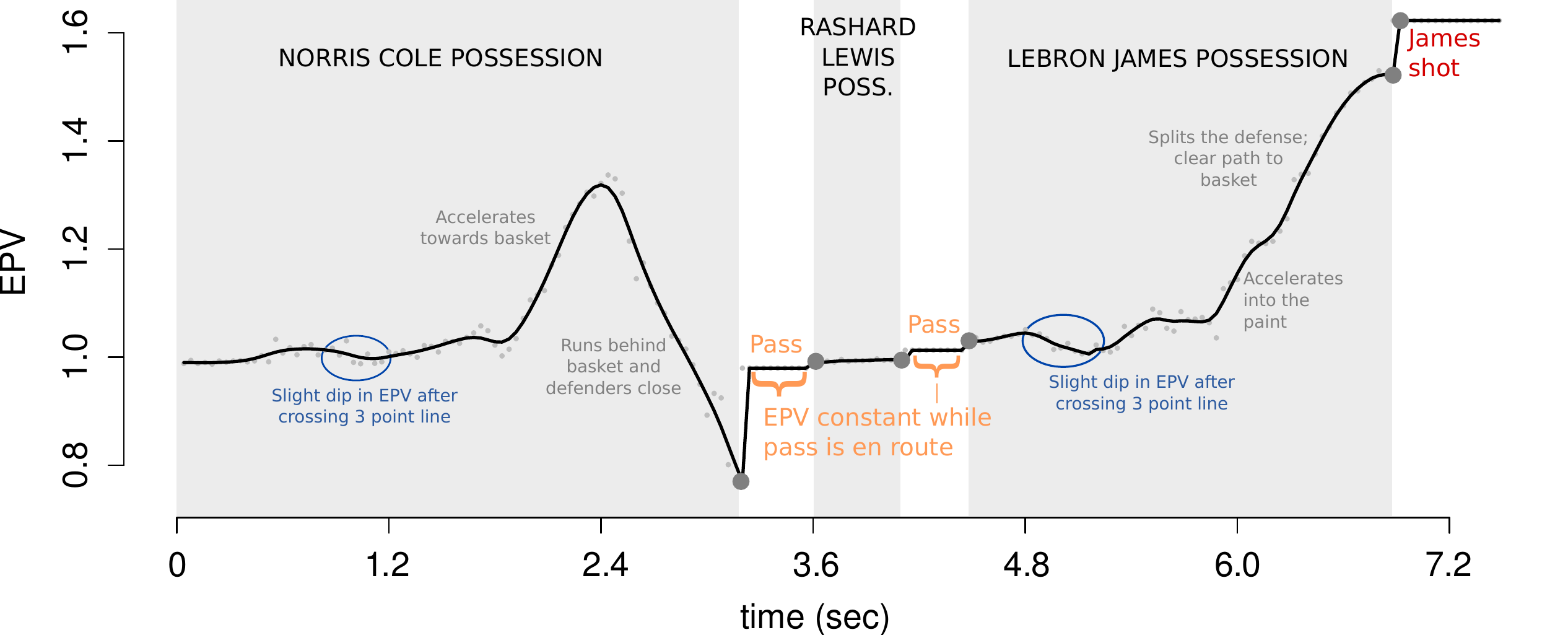}
\caption{Estimated EPV over time for the possession shown in Figure \ref{heat_poss}. Changes in EPV are induced by changes in players' locations and dynamics of motion; macrotransitions such as passes and shot attempts produce immediate, sometimes rapid changes in EPV. The black line slightly smooths EPV evaluations at each time point (gray dots), which are subject to Monte Carlo error.}
\label{heat_epv}
\end{figure}

For illustration, we plot the EPV curve corresponding to the example Heat possession in Figure \ref{heat_epv}, with EPV estimated using the methodology in this paper. We see several moments when the expected point yield of the possession, given its history, changes dramatically. For the first 2 seconds of the possession, EPV remains around 1. When Cole drives toward the basket, EPV rises until peaking at around 1.34 when Cole is right in front of the basket. As Cole dribbles past the basket (and his defenders continue pursuit), however, EPV falls rapidly, bottoming out at 0.77 before ``resetting'' to 1.00 with the pass to Rashard Lewis. The EPV increases slightly to 1.03 when the ball is then passed to James. As EPV is sensitive to small changes in players' exact locations, we see EPV rise slightly as James approaches the three point line and then dip slightly as he crosses it. Shortly afterwards, EPV rises suddenly as James breaks towards the basket, eluding the defense, and continues rising until he is beneath the basket, when an attempted layup boosts the EPV from 1.52 to 1.62.

In this way, EPV corresponds naturally to a coach's or spectator's sense of how the actions that basketball players take in continuous time help or hurt their team's cause to score in the current possession, and quantifies this in units of expected points.
EPV acts like a stock ticker, providing an instantaneous summary of the possession's eventual point value given all available information, much like a stock price values an asset based on speculation of future expectations.

\subsection{Related Work and Contributions}
Concepts similar to EPV, where final outcomes are modeled conditional on observed progress, have had statistical treatment in other sports, such as in-game win probability in baseball \cite{bukiet1997markov,yang2004two} and football \cite{lock2014using}, as well as in-possession point totals in football \cite{burke2010,goldner2012markov}. These previous efforts can be categorized into either marginal regression/classification approaches, where features of the current game state are mapped directly to expected outcomes, or process-based models that use a homogeneous Markov chain representation of the game to derive outcome distributions. Neither of these approaches is ideal for application to basketball. Marginal regression methodologies ignore the natural martingale structure of EPV, which is essential to its ``stock ticker'' interpretation. On the other hand, while Markov chain methodologies do maintain this ``stock ticker'' structure, applying them to basketball requires discretizing the data, introducing an onerous bias-variance-computation tradeoff that is not present for sports like baseball that are naturally discrete in time. 

To estimate EPV effectively, we introduce a novel multiresolution approach in which we model basketball possessions at two separate levels of resolution, one fully continuous and one highly coarsened. By coherently combining these models we are able to obtain EPV estimates that are reliable, sensitive, stochastically consistent, and computationally feasible (albeit intensive).
While our methodology is motivated by basketball, we believe that this research
can serve as an informative case study for analysts working in other application areas where continuous monitoring data are becoming widespread, including traffic monitoring \cite{ihler2006adaptive}, surveillance, and digital marketing \cite{shao2011data}, as well as other sports such as soccer and hockey \cite{thomas2013competing}. 

Section \ref{sec:Multiresolution} formally defines EPV within the context of a stochastic process for basketball, introducing the multiresolution modeling approach that make EPV calculations tractable as averages over future paths of a stochastic process. Parameters for these models, which represent players' decision-making tendencies in various spatial and situational circumstances, are discussed in Section \ref{sec:Macro}. Section \ref{sec:Computation} discusses inference for these parameters using hierchical models that share information between players and across space. We highlight results from actual NBA possessions in Section \ref{sec:Results}, and show how EPV provides useful new quantifications of player ability and skill. Section \ref{sec:Discussion} concludes with directions for further work.
%

%% file: multires2.tex
To present our process model for a basketball possession, we require some formalism. Let $\Omega$ represent the space of all possible basketball possessions in full detail, with $\omega \in \Omega$ describing the full path of a particular possession. For simplicity, we restrict our focus to possessions that do not include fouls\footnote{Our data include foul events, but do not specify the type or circumstances of the foul. There are several types of fouls and game situations for which fouls lead to free throws---for instance, shooting fouls, technical/flagrant fouls, clear path fouls, and fouls during the fouling team's ``bonus'' period; thus, modeling fouls presents additional complications relative to the other events we model in our EPV model. While drawing fouls can be a valuable and important part of team strategy, we omit modeling such behavior in this paper.}, so that each possession we consider results in 0, 2, or 3 points scored for the offense, denoted $X(\omega) \in \{0, 2, 3\}$. 
For any possession path $\omega$, we denote by $Z(\omega)$ the optical tracking time series generated by this possession so that $Z_t(\omega) \in \Z$, $t > 0$, is a ``snapshot'' of the tracking data exactly $t$ seconds from the start of the possession ($t=0$). $\Z$ is a high dimensional space that includes $(x,y)$ coordinates for all 10 players on the court, $(x,y,z)$ coordinates for the ball, summary information such as which players are on the court and what the game situation is (game location, score, time remaining, etc.), and event annotations that are observable in real time, such as a turnover occurring, a pass, or a shot being attempted and the result of that attempt.

Taking the intuitive view of $\Omega$ as a sample space of possession paths, we model $Z(\omega)$ as a stochastic process, and likewise, define $Z_t(\omega)$ for each $t > 0$ as a random variable in $\Z$. $Z(\omega)$ provides the natural filtration $\Fz_t = \sigma(\{Z_s^{-1}: 0 \leq s \leq t\})$, which represents all information available from the optical tracking data for the first $t$ seconds of a possession.
EPV is the expected value of the number of points scored for the possession ($X$) given all available data up to time $t$ ($\Fz_t$):

\begin{definition}
    The \textit{expected possession value}, or EPV, at time $t \geq 0$ during a possession is $\nu_t = \E[X|\Fz_t]$.
\end{definition}


The expectation $\E[X|\Fz_t]$ is an integral over the distribution of future paths the current possession can take. Letting $T(\omega)$ denote the time at which a possession following path $\omega$ ends\footnote{The time of a possession is bounded, even for pathological examples, by the 12-minute length of a quarter; yet we do not leverage this fact and simply assume that possession lenghts are almost surely finite.}, the possession's point total is a deterministic function of the full resolution data at this time, $X(\omega) = h(Z_{T(\omega)}(\omega))$. Thus, evaluating EPV amounts to integrating over the joint distribution of $(T, Z_T)$:
\begin{align} \label{epvdef}
\nu_t = \E[X|\Fz_t] &= \int_{\Omega} X(\omega) \prob(d\omega | \Fz_t) \nonumber \\
 &= \int_t^{\infty} \int_{\Z} h(z) \prob(Z_s = z | T=s, \Fz_t) \prob(T=s | \Fz_t) dz ds.
\end{align}
Note that we use probability notation $\prob(\cdot)$ somewhat heuristically, as $\prob(T=s|\Fz_t)$ is a density with respect to Lebesgue measure, while $Z_s$ mixes both discrete (annotations) and continuous (locations) components. We will also generally omit the dependence on $\omega$ when writing random variables, e.g., $Z_t$ instead of $Z_t(\omega)$.

\subsection{Estimator Criteria}
\label{subsec:multiCrit}
We have defined EPV in \eqref{epvdef} as an unobserved, theoretical quantity; one could thus imagine many different EPV estimators based on different models and/or information in the data. However, we believe that in order for EPV to achieve its full potential as a basis for high-resolution player and strategy evaluation, an EPV estimator should meet several criteria.

First, we require that the EPV estimator be stochastically consistent. Recognizing that EPV is simply a conditional expectation, it is tempting to estimate EPV using a regression or classification approach that maps features from $\Fz_t$ to an outcome space, $[0, 3]$ or $\{0, 2, 3\}$. Setting aside the fact that our data associate each possession outcome $X$ with process-valued inputs $Z$, and thus do not conform naturally to input/output structure of such models, such an approach cannot guarantee the estimator will have the (Kolmogorov) stochastic consistency inherent to theoretical EPV, which is essential to its ``stock ticker'' interpretation. Using a stochastically consistent EPV estimator guarantees that changes in the resulting EPV curve derive from players' on-court actions, rather than artifacts or inefficiencies of the data analysis.
A stochastic process model for the evolution of a basketball possession guarantees such consistency.

The second criterion that we require is that the estimator be sensitive to the fine-grained details of the data without incurring undue variance or computatonal complexity. Applying a Markov chain-based estimation approach would require discretizing the data by mapping the observed spatial configuration $Z_t$ into a simplified summary $C_t$, violating this criterion by trading potentially useful information in the player tracking data for computational tractability.

To develop methodology that meet both criteria, we note the information-computation tradeoff in current process modeling strategies results from choosing a single level of resolution at which to model the possession and compute all expectations. In contrast, our method for estimating EPV combines models for the possession at two distinct levels of resolution, namely, a fully continuous model of player movement and actions, and a Markov chain model for a highly coarsened view of the possession. This multiresolution approach leverages the computational simplicity of a discrete Markov chain model while conditioning on exact spatial locations and high-resolution data features.

\subsection{A Coarsened Process}\label{subsec:multiCoarse}
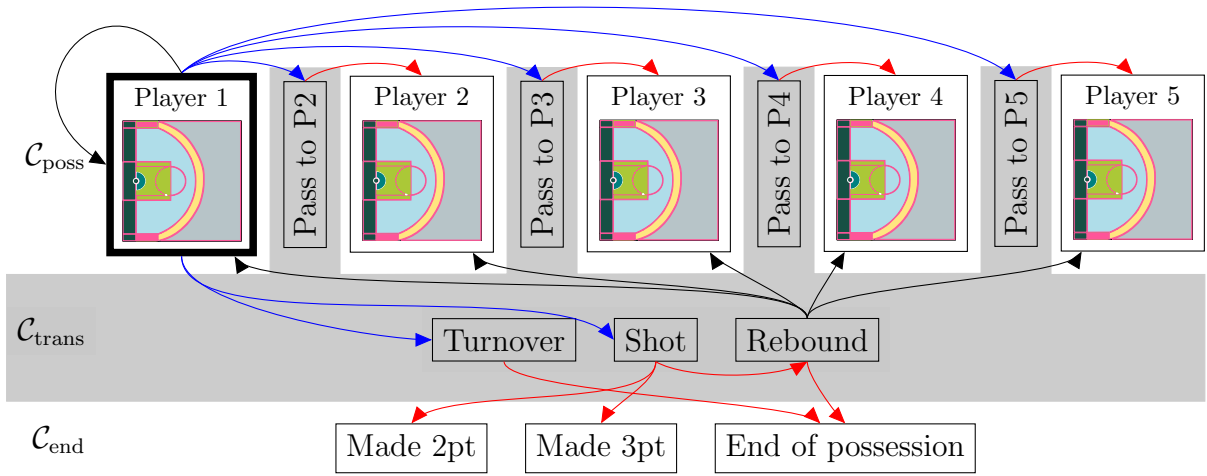
\begin{figure}[!ht]
\input{EPV_coarsened_tikz.tex}
\caption{Schematic of the coarsened possession process $C$, with states (rectangles) and possible state transitions (arrows) shown. The unshaded states in the first row compose $\Cset_{\text{poss}}$. Here, states corresponding to distinct ballhandlers are grouped together (Player 1 through 5), and the discretized court in each group represents the player's coarsened position and defended state. The gray shaded rectangles are transition states, $\Cset_{\text{trans}}$, while the rectangles in the third row represent the end states, $\Cset_{\text{end}}$. Blue arrows represent possible macrotransition entrances (and red arrows, macrotransition exits) when Player 1 has the ball; these terms are introduced in Section \ref{sec:Macro}.}
\label{fig:states}
\end{figure}

The Markov chain portion of our method requires a coarsened view of the data. For all time $0 < t \leq T$ during a possession, let $C(\cdot)$ be a coarsening that maps $\Z$ to a finite set $\Cset$, and call $C_t = C(Z_t)$ the ``state'' of the possession. To make the Markovian assumption plausible, we populate the coarsened state space $\Cset$ with summaries of the full resolution data so that transitions between these states represent meaningful events in a basketball possession---see Figure~\ref{fig:states} for an illustration.

First, there are 3 ``bookkeeping'' states, denoted $\Cset_{\text{end}}$, that categorize the end of the possession, so that $C_T \in \Cset_{\text{end}}$ and for all $t < T, C_t \not \in \Cset_{\text{end}}$ (shown in the bottom row of Figure~\ref{fig:states}). These are $\Cset_{\text{end}} = $\{made 2 pt, made 3 pt, end of possession\}.
These three states have associated point values of 2, 3, and 0, respectively (the generic possession end state can be reached by turnovers and defensive rebounds, which yield no points). This makes the possession point value $X$ a function of the final coarsened state $C_T$.

Next, whenever a player possesses the ball at time $t$, we assume $C_t = (\text{ballcarrier ID at }t) \times (\text{court region at }t) \times (\text{defended at }t)$, having defined seven disjoint regions of the court and classifying a player as defended at time $t$ by whether there is a defender within 5 feet of him. The possible values of $C_t$, if a player possesses the ball at time $t$, thus live in $\Cset_{\text{poss}} = \{\text{player ID}\} \times \{\text{region ID}\} \times \{\mathbf{1}[\text{defended}]\}$. These states are represented by the unshaded portion of the top row of Figure~\ref{fig:states}, where the differently colored regions of the court diagrams reveal the court space discretization.

Finally, we define a set of states to indicate that an annotated basketball action from the full resolution data $Z$ is currently in progress. These ``transition'' states encapsulate constrained motifs in a possession, for example, when the ball is in the air traveling between players in a pass attempt. Explicitly, denote $\Cset_{\text{trans}} = \{$shot attempt from $c \in \Cset_{\text{poss}}$, pass to $c' \in \Cset_{\text{poss}}$ from $c \in \Cset_{\text{poss}}$, turnover in progress, rebound in progress$\}$ (listed in the gray shaded portions of Figure~\ref{fig:states}). These transition states carry information about the possession path, such as the most recent ballcarrier, and the target of the pass, while the ball is in the air during shot attempts and passes\footnote{The reason we index transition states by the origin of the pass/shot attempt (and destination of the pass) is to preserve this information under a Markov assumption, where generic ``pass'' or ``shot'' states would inappropriately allow future states to be independent of the players involved in the shot or pass.}. Note that, by design, a possession must pass through a state in $\Cset_{\text{trans}}$ in order to reach a state in $\Cset_{\text{end}}$. For simplicity and due to limitations of the data, this construction of $\Cset = \Cset_{\text{poss}} \cup \Cset_{\text{trans}} \cup \Cset_{\text{end}}$ excludes several notable basketball events (such as fouls, violations, and other stoppages in play) and aggregates others (the data, for example, does not discriminate among steals, intercepted passes, or lost balls out of bounds).



\subsection{Combining Resolutions}\label{subsec:multiTheory}
We make several modeling assumptions about the processes $Z$ and $C$, which allow them to be combined into a coherent EPV estimator. 
\begin{enumerate}[label=(A\arabic*)]
    \item $C$ is marginally semi-Markov.\label{A1}
\end{enumerate}
The semi-Markov assumption \ref{A1} guarantees that the embedded sequence of disjoint possession states $C^{(0)}, C^{(1)}, \ldots, C^{(K)}$ is a Markov chain, which ensures that it is straightforward to compute $\E[X | C_t]$ using the associated transition probability matrix \cite{kemeny1976finite}.

Next, we specify the relationship between coarsened and full-resolution conditioning. This first requires defining two additional time points which mark changes in the future evolution of the possession:
\begin{align}
\tau_t & = \begin{cases}
\text{min} \{ s : s > t, C_s \in \Cset_{\text{trans}}\} & \text{if } C_t \in \Cset_{\text{poss}} \\
t & \text{if } C_t \not \in \Cset_{\text{poss}}
\end{cases} \label{taudef} \\
\delta_t &= \text{min}\{s : s \geq \tau_t, C_s \not \in \Cset_{\text{trans}} \}
\label{deltadef}.
\end{align}
Thus, assuming a player possesses the ball at time $t$, $\tau_t$ is the first time after $t$ he attempts a shot/pass or turns the ball over (entering a state in $\Cset_{\text{trans}}$), and $\delta_t$ is the endpoint of this shot/pass/turnover (leaving a state in $\Cset_{\text{trans}}$). We assume that passing through these transition states, $\Cset_{\text{trans}}$, \textit{decouples} the future of the possession after time $\delta_t$ with its history up to time $t$:
\begin{enumerate}[label=(A\arabic*),resume]
\item For all $s > \delta_t$ and $c \in \Cset$, $\prob(C_s =c | C_{\delta_t}, \Fz_t) = \prob(C_s =c | C_{\delta_t})$. \label{A2} 
\end{enumerate}

Intuitively, assumption \ref{A2} states that for predicting coarsened states beyond some point in the future $\delta_t$, all information in the possession history up to time $t$ is summarized by the distribution of $C_{\delta_t}$. The dynamics of basketball make this assumption reasonable; when a player passes the ball or attempts a shot, this represents a structural transition in the basketball possession to which all players react. Their actions prior to this transition are not likely to influence their actions after this transition. Given $C_{\delta_t}$---which, for a pass at $\tau_t$ includes the pass recipient, his court region, and defensive pressure, and for a shot attempt at $\tau_t$ includes the shot outcome---data prior to the pass/shot attempt are not informative of the possession's future evolution.


Together, these assumptions yield a simplified expression for \eqref{epvdef}, which combines contributions from full-resolution and coarsened views of the process.
\begin{theorem}\label{epvtheorem}
    Under assumptions \ref{A1}--\ref{A2}, the full-resolution EPV $\nu_t$ can be rewritten:
\begin{equation}\label{epveqn}
\nu_t = \sum_{c \in \Cset} \E[X | C_{\delta_t} = c]\prob(C_{\delta_t} = c | \Fz_t).
\end{equation}
\end{theorem}
\begin{remark}
Although we have specified this result in terms of the specific coarsening defined in Section~\ref{subsec:multiCoarse}, we could substitute any coarsening for which \ref{A1}--\ref{A2} are well-defined and reasonably hold. We briefly discuss potential alternative coarsenings in Section \ref{sec:Discussion}.
\end{remark}
The proof of Theorem~\ref{epvtheorem}, follows immediately from \ref{A1}--\ref{A2}, and is therefore omitted. Heuristically, \eqref{epveqn} expresses $\nu_t$ as the expectation given by a homogeneous Markov chain on $\Cset$ with a random starting point $C_{\delta_t}$, where only the starting point depends on the full-resolution information $\Fz_t$. This result illustrates the multiresolution conditioning scheme that makes our EPV approach computationally feasible: the term $\E[X | C_{\delta_t} = c]$ is easy to calculate using properties of Markov chains, and $\prob(C_{\delta_t} | \Fz_t)$ only requires forecasting the full-resolution data for a short period of time relative to \eqref{epvdef}, as $\delta_t \leq T$.

%% file: EPV_coarsened_tikz.tex
\begin{tikzpicture}[remember picture]

    \matrix (mainmat) [column 1/.style={anchor=base east},
            column 2/.style={anchor=center},
            row 2/.style={nodes={fill=black!21}},
            row sep=3ex, node distance=0, xshift=-9ex]{

     \node[classlab] (posslab) {$\Cset_{\text{poss}}$};&
    
     \node (R1) {
         \tikz{
     \node[court]    (C1) {\includecourt}; %
     \player[line width=1mm] {player1}{ %
       (C1)%
     }{Player 1};
     \node[pass, right=10pt of player1, anchor=north](to-p2){Pass to P2};
     \node[court,right=13pt of to-p2.south, yshift=-6pt]    (C2)      {\includecourt} ; %
     \player {player2} { %
       (C2)
     } {Player 2};%
     \node[pass, right=10pt of player2,anchor=north](to-p3){Pass to P3};
     \node[court, right=13pt of to-p3.south, yshift=-6pt]    (C3)      {\includecourt} ; %
     \player {player3} { %
       (C3)
     } {Player 3};%
     \node[pass, right=10pt of player3,anchor=north](to-p4){Pass to P4};
     \node[court, right=13pt of to-p4.south, yshift=-6pt]    (C4)      {\includecourt} ; %
     \player {player4} { %
       (C4)
     } {Player 4};%
     \node[pass, right=10pt of player4,anchor=north](to-p5){Pass to P5};
     \node[court, right=13pt of to-p5.south, yshift=-6pt]    (C5)      {\includecourt} ; %
     \player {player5} { %
       (C5)
     } {Player 5};
     }};\\%

     \node[classlab](macrolab){$\Cset_{\text{trans}}$};&

     \node(R2) {
         \tikz{
     \node[state]     (turnover) {Turnover};
     \node[state, right=of turnover]     (shot)     {Shot}; 
     \node[state, right=of shot]     (rebound)  {Rebound};
     }};\\

     \node[classlab] (endlab)  {$\Cset_{\textrm{end}}$}; &

     \node (R4) {
         \tikz{
     \node[state]    (2pts)  {Made 2pt};
     \node[state, right=of 2pts]    (3pts)  {Made 3pt};
     \node[state, right=of 3pts]    (0pts)  {End of possession};
     }};\\
    };

  \draw[->, blue] (player1.north) to [looseness=0.6] (to-p2.east);
  \draw[->, blue] (player1.north) to [looseness=0.4] (to-p3.east);
  \draw[->, blue] (player1.north) to [looseness=0.4] (to-p4.east);
  \draw[->, blue] (player1.north) to [looseness=0.4] (to-p5.east);

  \draw[->, blue] (player1.south) to [out=270,in=130, looseness=0.5] (shot.west);
  \draw[->, blue] (player1.south) to [out=270,in=180, looseness=0.5] (turnover.west);

  \draw[->, red](to-p2.east) to [looseness=0.6] (player2.north);
  \draw[->, red](to-p3.east) to [looseness=0.6] (player3.north);
  \draw[->, red](to-p4.east) to [looseness=0.6] (player4.north);
  \draw[->, red](to-p5.east) to [looseness=0.6] (player5.north);

  \draw[->](rebound.north) to [out=90, in=300, looseness=0.23] (player1);
  \draw[->](rebound.north) to [out=90, in=300, looseness=0.3] (player2);
  \draw[->](rebound.north) to [out=90, in=300, looseness=0.3] (player3);
  \draw[->](rebound.north) to [out=90, in=240, looseness=0.3] (player4);
  \draw[->](rebound.north) to [out=90, in=240, looseness=0.3] (player5);

  \draw[->, red](shot.south) to [bend right, looseness=0.6] (rebound.south);
  \draw[->, red](shot.south) to [out=270, in=50, looseness=0.5] (2pts.north);
  \draw[->, red](shot.south) to [out=270, in=60, looseness=0.7] (3pts.north);

  \draw[->, red](turnover.south) to [out=270, looseness=0.25] (0pts);
  \draw[->, red](rebound.south) to [out=270, looseness=0.5] (0pts.north);

  \path[->] (player1.north) edge[loop, looseness=3, out=120, in=150] (player1.west);

  \foreach \x in {2,...,5}
  {
      \coordinate [above left=5pt and 5pt of to-p\x.north east] (ne\x);
      \coordinate [below left=10pt and 5pt of to-p\x.north west] (nw\x);
      \coordinate [below right=10pt and 5pt of to-p\x.south west] (sw\x);
      \coordinate [above right=5pt and 5pt of to-p\x.south east] (se\x);
  }

  \coordinate [below=15pt of turnover] (bto);

  \begin{pgfonlayer}{background}
  \draw[draw=black!20, fill=black!20] (ne2) -- (ne2 |- sw2) --
    (macrolab.west |- sw2) -- (macrolab.west |- bto) -- (bto -| player5.east) --
    (sw2 -| player5.east) -- (sw2 -| sw5) -- 
             (se5) -- (ne5) -- (nw5) --
    (sw4) -- (se4) -- (ne4) -- (nw4) --
    (sw3) -- (se3) -- (ne3) -- (nw3) --
    (sw2) -- (se2) -- cycle; 
  \end{pgfonlayer}
\end{tikzpicture}

%% file: EPV_macro.tex
The representation in Theorem \eqref{epvtheorem} shows that estimating EPV does not require a full-blown model for entire basketball possessions at high resolution. Instead, the priority is to accurately predict the next major ``decoupling'' action in the possession, which we have denoted $C_{\delta_t}$. At this point, Equation \eqref{epveqn} switches resolutions: $C_{\delta_t}$ depends on the full-resolution possession history $\Fz_t$, after which our EPV estimate only depends on the coarsened state $C_{\delta_t}$. This section presents models that operate on these two distinct levels of resolution, using parameterizations that reflect players' reactions to the situational, spatiotemporal predicaments they face on the basketball court.

First, using the full-resolution data, we need to predict $C_{\delta_t}$. We achieve this using three models; heuristically speaking, one predicts player movement in space while the ballcarrier remains constant, one predicts the occurrence of events (passes/shots/turnovers) that change the ballcarrier, and one predicts the outcome state ($C_{\delta_t}$) of such events. 

Writing these models requires some additional notation.
Fix $\epsilon > 0$, and for any $t \geq 0$ during the possession (here, we use 1/25 second since this is the temporal resolution of our data), and let $M(t)$ be the event $\{\tau_t \leq t + \epsilon\}$; for this, we say that a \textit{macrotransition} occurs during $(t, t + \epsilon]$. Recalling the definition of $\tau_t$ \eqref{taudef}, $M(t)$ is realized when the possession moves from $\Cset_{\text{poss}}$ to $\Cset_{\text{trans}}$, which represents the start of a pass, shot attempt, or turnover (and $M(t)$ is continuously realized throughout the duration of this action). We now define:
\begin{enumerate}[label=(M\arabic*)]
    \item The \emph{microtransition model}, $\prob(Z_{t + \epsilon} | M(t)^c, \Fz_t)$, which describes infinitesimal player movement assuming that a major ball movement does \emph{not} occur. \label{M3}
    \item The \emph{macrotransition entry model}, $\prob(M(t) | \Fz_t)$, which describes the occurrence of a macrotransition (pass/shot/turnover) within the next $\epsilon$ time.  \label{M1}
    \item The \emph{macrotransition exit model} $\prob(C_{\delta_t} | M(t), \Fz_t)$, which gives the outcome of this macrotransition in $\Cset$. \label{M2}
\end{enumerate}

\ref{M3}--\ref{M2} together allow us to sample from $\prob(C_{\delta_t} | \Fz_t)$, as we alternate draws from \ref{M3} and \ref{M1} until a macrotransition occurs, and then use \ref{M2} to sample the outcome state of this macrotransition. Thus, while models \ref{M3}--\ref{M2} condition on the full-resolution possession history $\Fz_t$ in order to predict $C_{\delta_t}$, calculating EPV given $C_{\delta_t}$ only requires a model for transitions between coarsened states $C$. Due to the Markov assumption \ref{A1}, this is easily summarized by a transition probability matrix:
\begin{enumerate}[label=(M\arabic*)]
        \setcounter{enumi}{3}
    \item The \emph{Markov transition probability matrix} $\mathbf{P}$, with $P_{qr} = \prob(C^{(n+1)} = c_r | C^{(n)} = c_q)$. \label{M4}
\end{enumerate}
Thus, \ref{M3}--\ref{M4} are sufficient to compute EPV using our multiresolution framework of Theorem \ref{epvtheorem}. In the following subsections, we discuss each of these models in greater detail.

\subsection{Microtransition Model}

The microtransition model describes player movement with the ballcarrier held constant. In the periods between transfers of ball possession (including passes, shots, and turnovers), all players on the court move in order to influence the character of the next ball movement (macrotransition). For instance, the ballcarrier might drive toward the basket to attempt a shot, or move laterally to gain separation from a defender, while his teammates move to position themselves for passes or rebounds, or to set screens and picks. The defense moves correspondingly, attempting to deter easy shot attempts or passes to certain players while simultaneously anticipating a possible turnover.
Separate models are assumed for offensive and defensive players, as we shall describe.

\begin{figure}[h]
\centering
\begin{tabular}{cccc}
\subfloat[]{\includegraphics[width=0.23\linewidth]{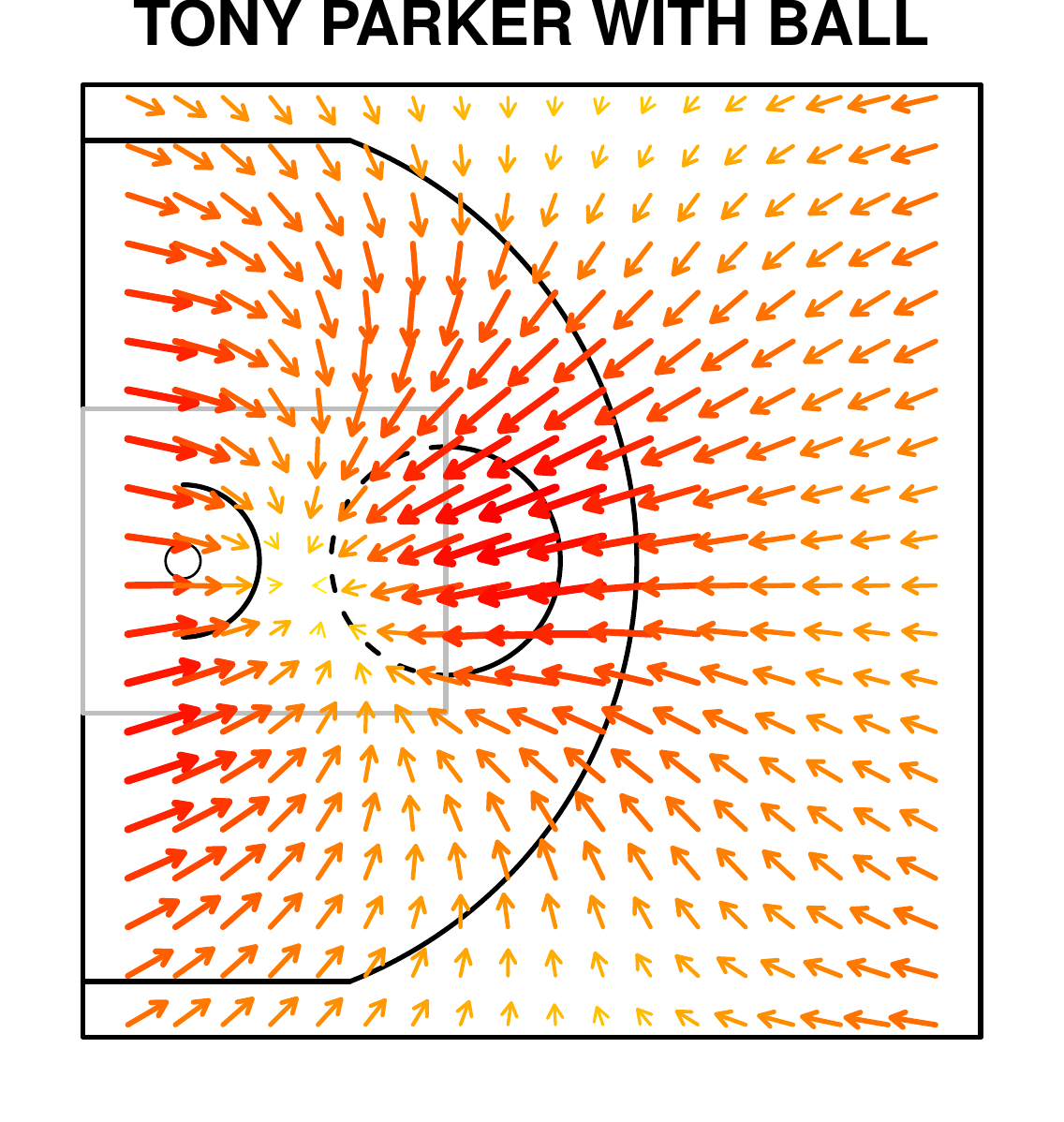}} &
\subfloat[]{\includegraphics[width=0.23\linewidth]{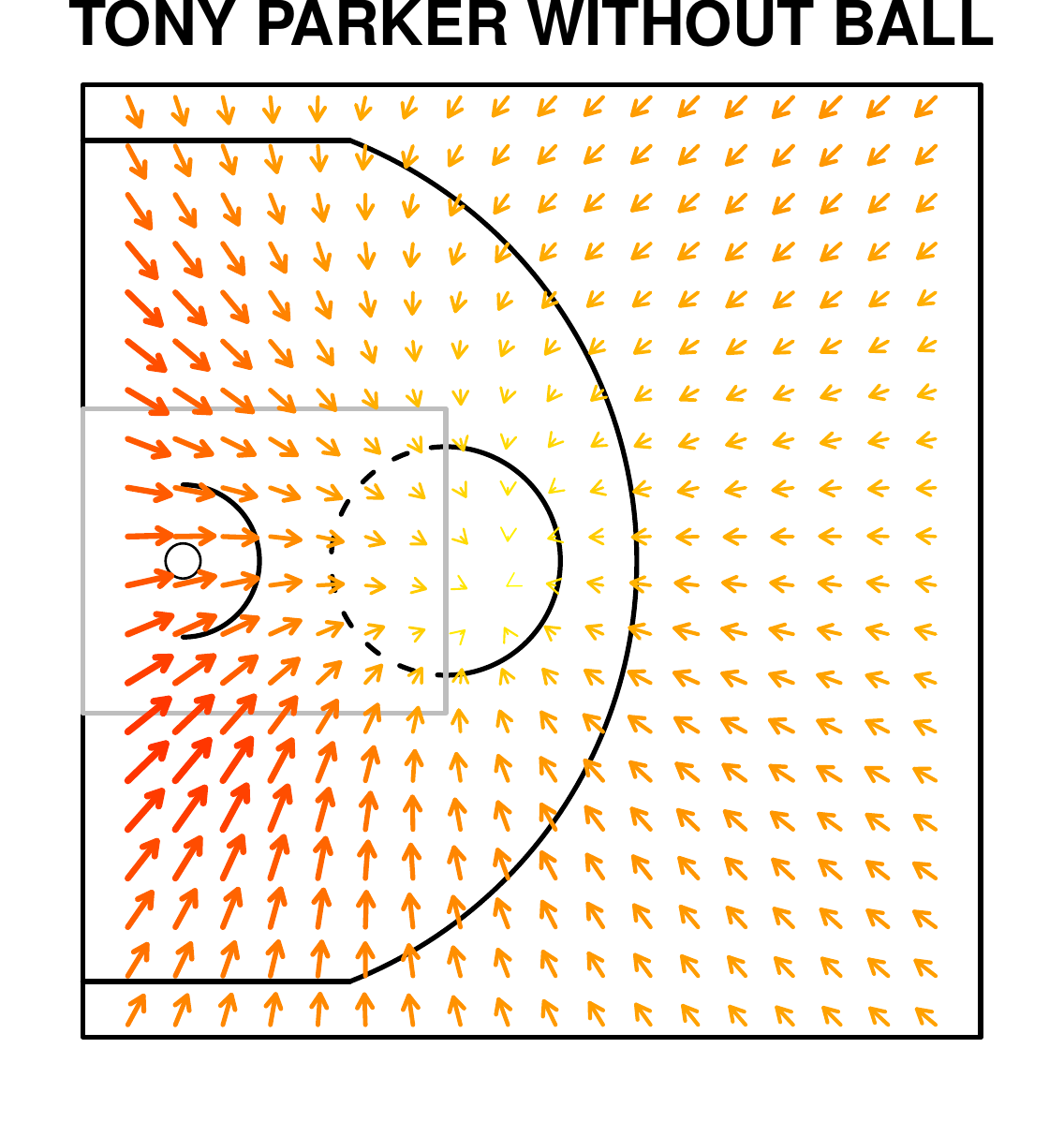}} &
\subfloat[]{\includegraphics[width=0.23\linewidth]{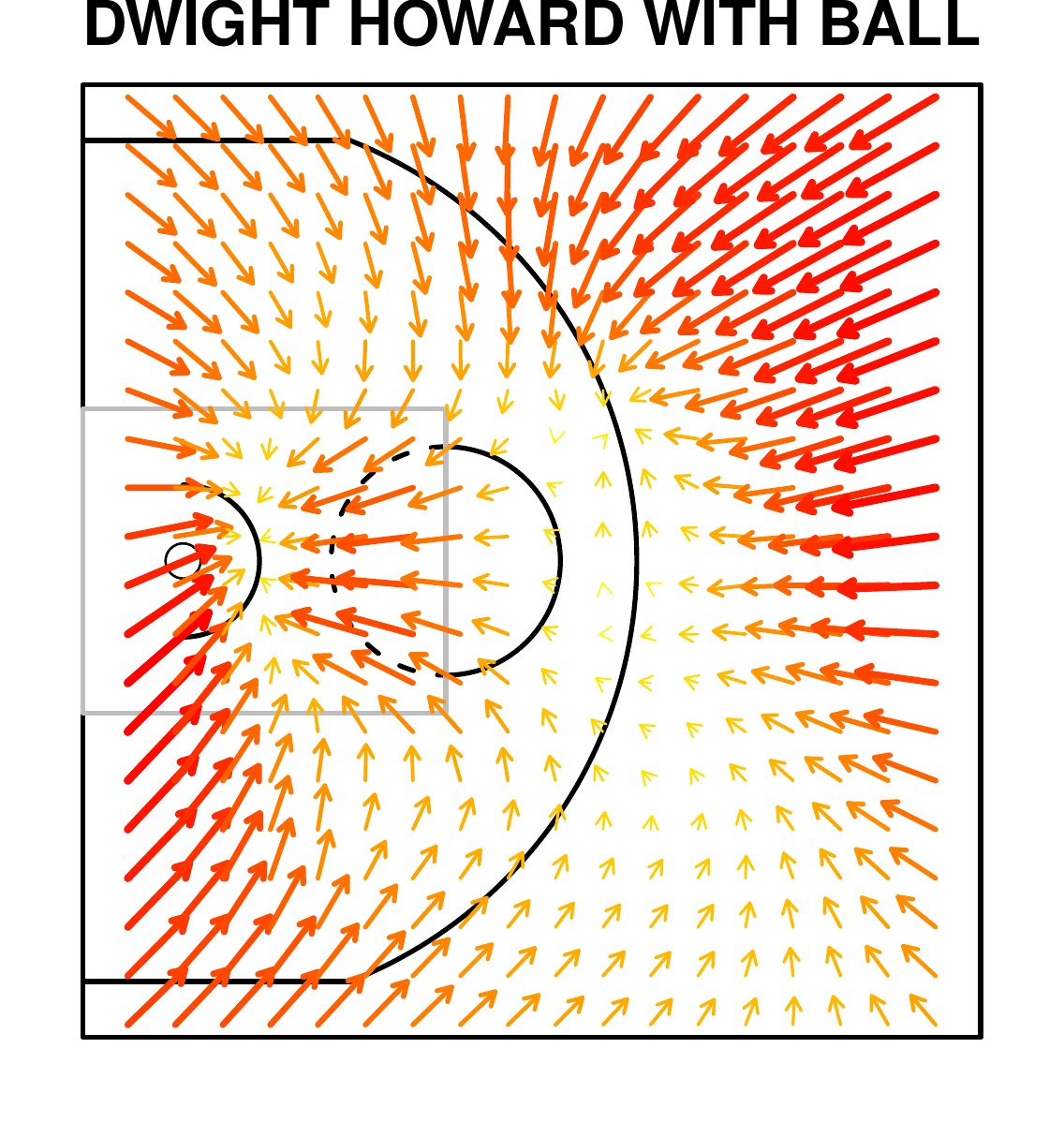}} &
\subfloat[]{\includegraphics[width=0.23\linewidth]{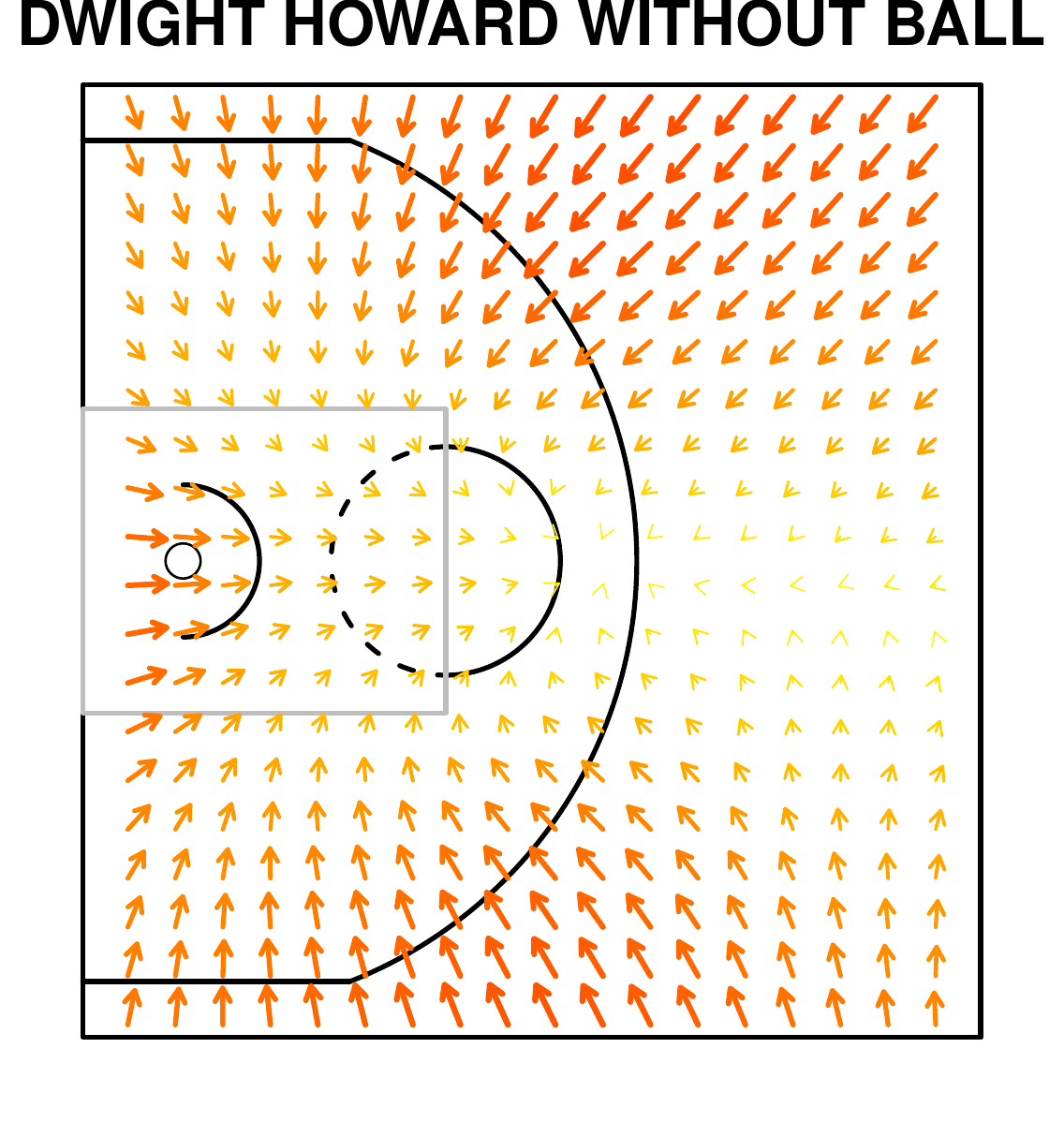}}
\end{tabular}
\caption{Acceleration fields $(\mu_x(\bz(t)), \mu_y(\bz(t)))$ for Tony Paker (a)--(b) and Dwight Howard (c)--(d) with and without ball possession. The arrows point in the direction of the acceleration at each point on the court's surface, and the size and color of the arrows are proportional to the magnitude of the acceleration. Comparing (a) and (c) for instance, we see that when both players possess the ball, Parker more frequently attacks the basket from outside the perimeter. Howard does not accelerate to the basket from beyond the perimeter, and only tends to attack the basket inside the paint.}
\label{accelerations}
\end{figure}

Predicting the motion of offensive players over a short time window is driven by the players' dynamics (velocity, acceleration, etc.).  Let the location of offensive player $\ell$ ($\ell \in \{1, \ldots, L = 461\}$) at time $t$ be $\mathbf{z}^{\ell}(t) = (x^{\ell}(t), y^{\ell}(t))$. We then model movement in each of the $x$ and $y$ coordinates using
\begin{align}
x^{\ell}(t + \epsilon) &=  x^{\ell}(t) + \alpha^{\ell}_x[x^{\ell}(t) - x^{\ell}(t - \epsilon)] + \eta^{\ell}_x(t)
\label{micro}
\end{align}
(and analogously for $y^{\ell}(t)$). This expression derives from a Taylor series expansion to the ballcarrier's position for each coordinate, such that $\alpha^{\ell}_x[x^{\ell}(t) - x^{\ell}(t - \epsilon)] \approx \epsilon x^{\ell \prime}(t)$, and $\eta^{\ell}_x(t)$ provides stochastic innovations representing the contribution of higher-order derivatives (acceleration, jerk, etc.).
Because they are driven to score, players' dynamics on offense are nonstationary. When possessing the ball, most players accelerate toward the basket when beyond the three-point line, and decelerate when very close to the basket in order to attempt a shot.  Also, players will accelerate away from the edges of the court as they approach these, in order to stay in bounds. To capture such behavior, we assume spatial structure for the innovations, $\eta^{\ell}_x(t) \sim \norm (\mu^{\ell}_x(\mathbf{z}^{\ell}(t)), (\sigma^{\ell}_x)^2)$, where $\mu^{\ell}_x$ maps player $\ell$'s location on the court to an additive effect in \eqref{micro}, which has the interpretation of an acceleration effect; see Figure \ref{accelerations} for an example.


The defensive components of $\prob(Z_{t+\epsilon} | M(t)^c, \Fz_t)$, corresponding to the positions of the five defenders, are easier to model conditional on the evolution of the offense's positions. Following \citeasnoun{franks2014defensive}, we assume each defender's position is centered on a linear combination of the basket's location, the ball's location, and the location of the offensive player he is guarding. \citeasnoun{franks2014defensive} use a hidden Markov model (HMM) based on this assumption to learn which offensive players each defender is guarding, such that conditional on defender $\ell$ guarding offender $k$ his location $\bz^{\ell}(t) = (x^{\ell}(t), y^{\ell}(t))$ should be normally distributed with mean $\mathbf{m}^k_{\text{opt}}(t) = 0.62\bz^k(t) + 0.11\bz_{\text{bask}} + 0.27\bz_{\text{ball}}(t)$. 

Of course, the dynamics (velocity, etc.) of defensive players' are still hugely informative for predicting their locations within a small time window. Thus our microtransition model for defender $\ell$ balances these dynamics with the mean path induced by the player he is guarding:
\begin{align}
x^{\ell}(t + \epsilon)|m^k_{\text{opt}, x}(t) & \sim  \N \bigg( x^{\ell}(t) + a^{\ell}_x[x^{\ell}(t) - x^{\ell}(t-\epsilon)] \nonumber \\
& \hspace{-2cm} + b^{\ell}_x[m^k_{\text{opt}, x}(t + \epsilon) - m^k_{\text{opt}, x}(t)] + c_x^{\ell}[x^{\ell}(t) - m^k_{\text{opt}, x}(t + \epsilon)], (\tau^{\ell}_x)^2\bigg)
\label{micro-defense}
\end{align}
and symmetrically in $y$. Rather than implement the HMM procedure used in \citeasnoun{franks2014defensive}, we simply assume each defender is guarding at time $t$ whichever offensive player $j$ yields the smallest residual $||\bz^{\ell}(t)  - \mathbf{m}^j_{\text{opt}}(t)||$, noting that more than one defender may be guarding the same offender (as in a ``double team'').  Thus, conditional on the locations of the offense at time $t+\epsilon$, \eqref{micro-defense} provides a distribution over the locations of the defense at $t + \epsilon$. 

\subsection{Macrotransition Entry Model}

The macrotransition entry model $\prob(M(t) | \Fz_t)$ predicts ball movements that instantaneously shift the course of the possession---passes, shot attempts, and turnovers. As such, 
we consider a family of macrotransition entry models $\prob(M_j(t)
|\Fz_t)$, where $j$ indexes the type of macrotransition corresponding
to $M(t)$. There are six such types: four pass options (indexed,
without loss of generality, $j \leq 4$), a shot attempt ($j = 5$), or
a turnover $(j=6)$. Thus, $M_j(t)$ is the event that a macrotransition
of type $j$ begins in the time window $(t, t + \epsilon]$, and $M(t) =
\bigcup_{j=1}^6 M_j(t)$. Since macrotransition types are disjoint, we
also know $\prob(M(t) | \Fz_t) = \sum_{j=1}^6 \prob(M_j(t) | \Fz_t)$.

We parameterize the macrotransition entry models as competing risks \cite{prentice1978analysis}: assuming player $\ell$ possesses the ball at time $t > 0$ during a possession, denote
\begin{equation}\label{hazard-def}
\lambda^{\ell}_j (t) = \lim_{\epsilon \rightarrow 0} \frac{\prob(M_j(t) |\Fz_t)}{\epsilon}
\end{equation}
as the hazard for macrotransition $j$ at time $t$.
We assume these are log-linear,
\begin{equation}\label{hazard-equation}
\log(\lambda^{\ell}_j(t)) = [\mathbf{W}_j^{\ell}(t)]'\boldsymbol{\beta}_j^{\ell} + \xi_j^{\ell}\left(\bz^{\ell}(t)\right) + \left(\tilde{\xi}_j^{\ell}\left(\bz_{j}(t)\right)\mathbf{1}[j \leq 4]\right),
\end{equation}
where $\mathbf{W}_j^{\ell}(t)$ is a $p_j \times 1$ vector of time-varying covariates, $\boldsymbol{\beta}_j^{\ell}$ a $p_j \times 1$ vector of coefficients, $\bz^{\ell}(t)$ is the ballcarrier's 2D location on the court (denote the court space $\Ss$) at time $t$, and $\xi_j^{\ell}: \Ss \rightarrow \R$ is a mapping of the player's court location to an additive effect on the log-hazard, providing spatial variation. The last term in \eqref{hazard-equation} only appears for pass events $(j \leq 4)$ to incorporate the location of the receiving player for the corresponding pass: $\bz_j(t)$ (which slightly abuses notation) provides his location on the court at time $t$, and $\tilde{\xi}_j^{\ell}$, analogously to $\xi_j^{\ell}$, maps this location to an additive effect on the log-hazard. The four different passing options are identified by the (basketball) position of the potential pass recipient: point guard (PG), shooting guard (SG), small forward (SF), power forward (PF), and center (C). 

\begin{figure}
\centering
\begin{tabular}{cccc}
\subfloat[$\xi_1, \tilde{\xi}_1$ (pass to PG)]{\includegraphics[width=0.23\linewidth,height=0.08\textheight]{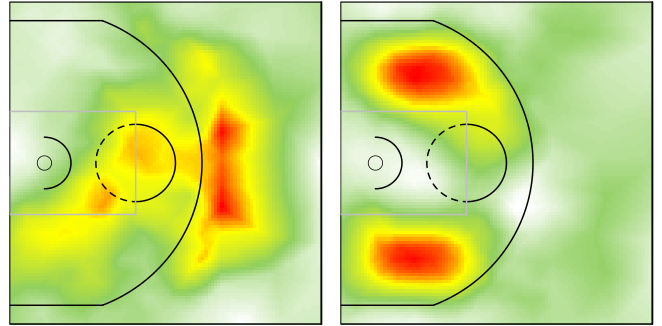}} & 
\subfloat[$\xi_2, \tilde{\xi}_2$ (pass to SG)]{\includegraphics[width=0.23\linewidth,height=0.08\textheight]{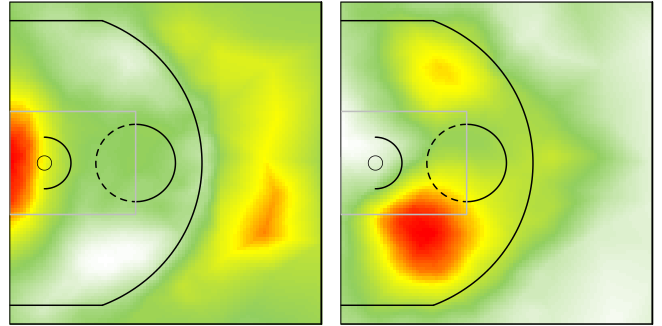}} &
\multirow{-2}[8]{*}{\subfloat[$\xi_5$ (shot-taking)]{\includegraphics[width=0.23\linewidth,height=0.17\textheight]{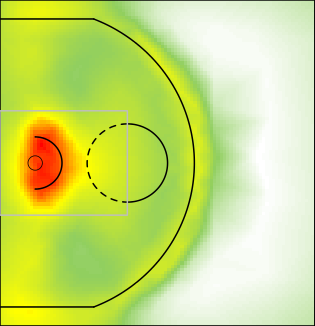}}} &
\multirow{-2}[8]{*}{\subfloat[$\xi_6$ (turnover)]{\includegraphics[width=0.23\linewidth,height=0.17\textheight]{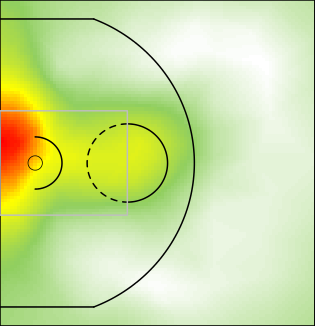}}}
\\
\subfloat[$\xi_3, \tilde{\xi}_3$ (pass to PF)]{\includegraphics[width=0.23\linewidth,height=0.08\textheight]{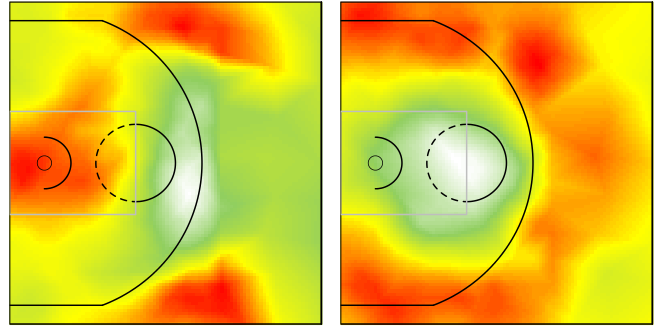}} &
\subfloat[$\xi_4, \tilde{\xi}_4$ (pass to C)]{\includegraphics[width=0.23\linewidth,height=0.08\textheight]{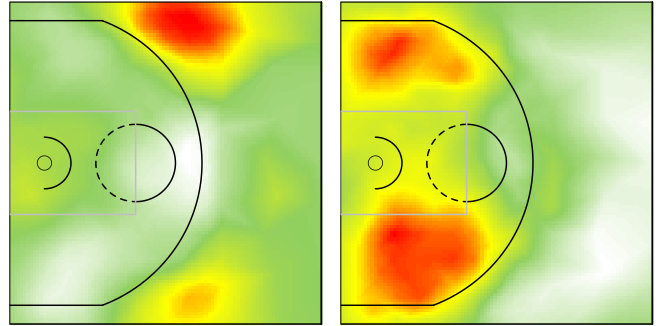}}
& &
\end{tabular}
\caption[Sample plots of spatial effects]{Plots of estimated spatial effects $\xi$ for LeBron James as the ballcarrier. For instance, plot (c) reveals the largest effect on James' shot-taking hazard occurs near the basket, with noticeable positive effects also around the three point line (particularly in the ``corner 3'' shot areas). Plot (a) shows that he is more likely (per unit time) to pass to the point guard when at the top of the arc---more so when the point guard is positioned in the post area.}
\label{fig:spatial_effects}
\end{figure}

The macrotransition model \eqref{hazard-def}--\eqref{hazard-equation} represents the ballcarrier's decision-making process as an interpretable function of the unique basketball predicaments he faces. For example, in considering the hazard of a shot attempt, the time-varying covariates ($\mathbf{W}_j^{\ell}(t)$) we use are the distance between the ballcarrier and his nearest defender (transformed as $\log(1+d)$ to moderate the influence of extremely large or small observed distances), an indicator for whether the ballcarrier has dribbled since gaining possession, and a constant representing a baseline shooting rate (this is not time-varying)\footnote{Full details on all covariates used for all macrotransition types are included in Appendix \ref{Covariates}}. The spatial effects $\xi_j^{\ell}$ reveal locations where player $\ell$ is more/less likely to attempt a shot in a small time window, holding fixed the time-varying covariates $\mathbf{W}_j^{\ell}(t)$. Such spatial effects (illustrated in Figure \ref{fig:spatial_effects}) are well-known to be nonlinear in distance from the basket and asymmetric about the angle to the basket \cite{miller2013icml}. 

%
All model components---the time-varying covariates, their coefficients, and the spatial effects $\xi, \tilde{\xi}$ differ across macrotransition types $j$ for the same ballcarrier $\ell$, as well as across all $L=461$ ballcarriers in the league during the 2013-14 season. This reflects the fact that players' decision-making tendencies and skills are unique; a player such as Dwight Howard will very rarely attempt a three point shot even if he is completely undefended, while someone like Stephen Curry will attempt a three point shot even when closely defended.


\subsection{Macrotransition Exit Model}
\label{macro_exit}
Using the six macrotransition types introduced in the previous subsection, we can express the macrotransition exit model \ref{M2} when player $\ell$ has possession as
\begin{align}
\prob(C_{\delta_t} | M(t), \Fz_t) &= \sum_{j=1}^6 \prob(C_{\delta_t} | M_j(t), \Fz_t) \prob(M_j(t) | M(t), \Fz_t) \nonumber \\
&= \sum_{j=1}^6 \prob(C_{\delta_t} |  M_j(t), \Fz_t) \frac{\lambda_j^{\ell}(t)}{\sum_{k=1}^6 \lambda^{\ell}_k(t)},
\label{exitmodel}
\end{align}
using the competing risks model for $M_j(t)$ given by \eqref{hazard-def}--\eqref{hazard-equation}. As terms $\lambda^{\ell}_j(t)$ are supplied by \eqref{hazard-equation}, we focus on the macrotransition exit model conditional on the macrotransition type, $\prob(C_{\delta_t} | M_j(t), \Fz_t)$.

For each $j=1, \ldots, 4$, $M_j(t)$ represents a pass-type macrotransition, therefore $C_{\delta_t}$ is a possession state $c' \in \Cset_{\text{poss}}$ for the player corresponding to pass option $j$. Thus, a model for $\prob(C_{\delta_t} | M_j(t), \Fz_t)$ requires us to predict the state $c' \in \Cset_{\text{poss}}$ the $j$th pass target will occupy upon receiving the ball. Our approach is to simply assume $c'$ is given by the pass target's location at the time the pass begins. While this is naive and could be improved by further modeling, it is a reasonable approximation in practice, because with only seven court regions and two defensive spacings comprising $\Cset_{\text{poss}}$, the pass recipient's position in this space is unlikely to change during the time the pass is traveling en route, $\delta_t - t$ (a noteable exception is the alley-oop pass, which leads the pass recipient from well outside the basket to a dunk or layup within the restricted area). Our approach thus collapses $\prob(C_{\delta_t} | M_j(t), \Fz_t)$ to a single state in $\Cset_{\text{poss}}$, which corresponds to pass target $j$'s location at time $t$.

When $j=5$, and a shot attempt occurs in $(t, t + \epsilon]$, $C_{\delta_t}$ is either a made/missed 2 point shot, or made/missed three point shot. For sufficiently small $\epsilon$, we observe at $Z_t$ whether the shot attempt in $(t, t + \epsilon]$ is a two- or three-point shot, therefore our task in providing $\prob(C_{\delta_t} | M_j(t), \Fz_t)$ is modeling the shot's probability of success. We provide a parametric shot probability model, which shares the same form as the macrotransition entry model \eqref{hazard-def}--\eqref{hazard-equation}, though we use a logit link function as we are modeling a probability instead of a hazard. Specifically, for player $\ell$ attempting a shot at time $t$, let $p^{\ell}(t)$ represent the probability of the shot attempt being successful (resulting in a basket). We assume
\begin{equation}\label{shotprob}
\text{logit}(p^{\ell}(t)) = [\mathbf{W}_\shot^{\ell}(t)]'\boldsymbol{\beta}_\shot^{\ell} + \xi_\shot^{\ell}(\bz^{\ell}(t))
\end{equation}
with components in \eqref{shotprob} having the same interpretation as their $j$-indexed counterparts in the competing risks model \eqref{hazard-equation}; that is, $\mathbf{W}_\shot^{\ell}$ is a vector of time-varying covariates (we use distance to the nearest defender---transformed as $\log(1 + d)$---an indicator for whether the player has dribbled, and a constant to capture baseline shooting efficiency) with $\boldsymbol{\beta}_\shot^{\ell}$ a corresponding vector of coefficients, and $\xi_\shot^{\ell}$ a smooth spatial effect, as in \eqref{hazard-equation}.

Lastly, when $j=6$ and $M_j(t)$ represents a turnover, $C_{\delta_t}$ is equal to the turnover state in $\Cset_{\text{end}}$ with probability 1. 

Note that the macrotransition exit model is mostly trivial when no player has ball possession at time $t$, since this implies $C_t \in \Cset_{\text{trans}} \cup \Cset_{\text{end}}$ and $\tau_t = t$. If $C_t \in \Cset_{\text{end}}$, then the possession is over and $C_{\delta_t} = C_t$. Otherwise, if $C_t \in \Cset_{\text{trans}}$ represents a pass attempt or turnover in progress, the following state $C_{\delta_t}$ is deterministic given $C_t$ (recall that the pass recipient and his location are encoded in the definition of pass attempt states in $\Cset_{\text{trans}}$). When $C_t$ represents a shot attempt in progress, the macrotransition exit model reduces to the shot probability model \eqref{shotprob}. Finally, when $C_t$ is a rebound in progress, we ignore full-resolution information and simply use the Markov transition probabilities from $\mathbf{P}$\footnote{Our rebounding model could be improved by using full-resolution spatiotemporal information, as players' reactions to the missed shot event are informative of who obtains the rebound.}.

\subsection{Transition Probability Matrix for Coarsened Process}


The last model necessary for calculating EPV is \ref{M4}, the
transition probability matrix for the embedded Markov chain
corresponding to the coarsened process $C^{(0)}, C^{(2)}, \ldots,
C^{(K)}$. This transition probability matrix is used to compute the
term $\E[X | C_{\delta_t} = c]$ that appears in Theorem
\ref{epvtheorem}. Recall that we denote the transition probability
matrix as $\mathbf{P}$, where $P_{qr} = \prob(C^{(i+1)} = c_r |
C^{(i)} = c_q)$ for any $c_q, c_r \in \Cset$.

Without any other probabilistic structure assumed for $C^{(i)}$ other
than Markov, for all $q,r$, the maximum likelihood estimator of
$P_{qr}$ is the observed transition frequency, 
$\hat{P}_{qr} = \frac{N_{qr}}{\sum_{r'}N_{qr'}}$, where $N_{qr}$ counts the number of
transitions $c_q \rightarrow c_r$. Of course, this estimator has
undesirable performance if the number of visits to any particular
state $c_q$ is small (for instance, Dwight Howard closely defended in the corner 3 region), as the estimated transition probabilities from
that state may be degenerate.  

Under our multiresolution model for basketball possessions, however,
expected transition counts between many coarsened states $C^{(i)}$ can
be computed as summaries of our macrotransition models \ref{M1}--\ref{M2}. To show this,
for any arbitrary $t > 0$ let $M_j^r(t)$ be the event
$$M_j^r(t) = \{\prob(M_j(t) \text{ and } C_{t + \epsilon} = c_r | \Fz_t) > 0\}.$$
Thus $M_j^r(t)$ occurs if it is possible for a macrotransition of type
$j$ into state $c_r$ to occur in $(t, t + \epsilon]$. When applicable, we can use this to get the expected number of $c_q
\rightarrow c_r$ transitions:
\begin{equation}
  \label{eq:shrunken_tprob}
\tilde{N}_{qr} = \epsilon \sum_{t : C_t = c_q} \lambda^{\ell}_j(t)\mathbf{1}[M_j^r(t)].
\end{equation}
When $c_q$ is a shot attempt state from $c_{q'} \in \Cset_{\text{poss}}$, \eqref{eq:shrunken_tprob} is
adjusted using the shot probability model \eqref{shotprob}:
$\tilde{N}_{qr} = \epsilon \sum_{t : C_t = c_{q'}}
\lambda^{\ell}_j(t)p(t)\mathbf{1}[M_j^r(t)]$ when $c_r$ represents an
eventual made shot and $\tilde{N}_{qr} = \epsilon \sum_{t : C_t = c_{q'}}
\lambda^{\ell}_j(t)(1 - p(t))\mathbf{1}[M_j^r(t)]$ when $c_r$ represents an
eventual miss. 

By replacing raw counts with their expectations conditional on higher-resolution data, leveraging the hazards $\eqref{eq:shrunken_tprob}$ provides a Rao-Blackwellized (unbiased, lower variance) alternative to counting observed transitions. Furthermore, due to the hierarchical parameterization of hazards $\lambda_j^{\ell}(t)$ (discussed in Section \ref{sec:Computation}), information is shared across space and player combinations so that estimated hazards are well-behaved even in situations without significant observed data. Thus, when $c_q \rightarrow c_r$ represents a macrotransition, we use $\tilde{N}_{qr}$ in place of $N_{qr}$ when calculating $\hat{P}_{qr}$.

%% file: EPV_computation.tex
A critical aspect of the micro- and macrotransition models defined in the previous section is that they are parameterized to capture the variations between actions, players, and court space that play a central role in basketball strategy. This section outlines the procedure for reliably estimating this rich set of model parameters using likelihood-based methods.

Hierarchical models are essential for our problem because by implicitly averaging over all possible future possession paths, calculating EPV requires transition probabilities for situations for which there is no data. For instance, DeAndre Jordan did not attempt a three-point shot in the 2013-14 season, yet any EPV estimate for a possession with him on the court requires an estimate of his shooting ability from everywhere on the court, even though for some of these regions it is unlikely he would attempt a shot. Hierarchical models combine information both across space and across different players to estimate such probabilities.


\subsection{Conditional Autoregressive Prior for Player-Specific Coefficients}
\label{subsec:CAR}
Sharing information between players is critical for our estimation problem, but standard hierarchical models encode an assumption of exchangeability between units that is too strong for NBA players, even between those who are classified by the league as playing the same position. For instance, LeBron James is listed at the same position (small forward) as Steve Novak, despite the fact that James is one of the NBA's most prolific short-range scorers whereas Novak has not scored a layup since 2012. To model between-player variation more realistically, our hierarchical model shares information across players based on a localized notion of player similarity that we represent as an $L \times L$ binary adjacency matrix $\mathbf{H}$: $H_{\ell k} = 1$ if players $\ell$ and $k$ are similar to each other and $H_{\ell k} = 0$ otherwise. We determine similarity in a pre-processing step that compares the spatial distribution of where players spend time on the offensive half-court; see Appendix \ref{subsec:H} for exact details on specifying $\mathbf{H}$. 

Now let $\beta^{\ell}_{ji}$ be the $i$th component of $\boldsymbol{\beta}^{\ell}_j$, the vector of coefficients for the time-referenced covariates for player $\ell$'s hazard $j$ \eqref{hazard-equation}. Also let $\boldsymbol{\beta}_{ji}$ be the vector representing this component across all $L = 461$ players, $(\beta^{1}_{ji} \: \: \beta^{2}_{ji} \: \ldots \: \beta^{L}_{ji})'$. We assume independent conditional autogressive (CAR) priors \cite{besag1974spatial} for $\boldsymbol{\beta}_{ji}$:
\begin{align}
\beta^{\ell}_{ji} | \beta^{(-\ell)}_{ji}, \tau_{\boldsymbol{\beta}_{ji}}^2 &\sim \norm \left( \frac{1}{n_{\ell}} \sum_{k : H_{\ell k} = 1} \beta^{k}_{ji}, \frac{\tau_{\boldsymbol{\beta}_{ji}}^2}{n_{\ell}} \right) \nonumber \\
\tau^2_{\boldsymbol{\beta}_{ji}} &\sim \text{InvGam}(1, 1)
\label{car}
\end{align}
where $n_{\ell} = \sum_{k} H_{\ell k}$. Similarly, let $\boldsymbol{\beta}_{\shot i} = (\beta^1_{\shot i} \: \: \beta^2_{\shot i} \: \ldots \: \beta^L_{\shot i})$ be the vector of the $i$th component of the shot probability model \eqref{shotprob} across players $1, \ldots, L$. We assume the same CAR prior \eqref{car} independently for each component $i$. While the inverse gamma prior for $\tau^2_{*}$ terms seems very informative, we want to avoid very large or small values of $\tau^2_{*}$, corresponding to 0 or full shrinkage (respectively), which we know are inappropriate for our model. Predictive performance for the 0 shrinkage model ($\tau^2_{*}$ very large) is shown in Table \ref{loglik-table}, whereas the full shrinkage model ($\tau^2_{*} = 0$) doesn't allow parameters to differ by player identity, which precludes many of the inferences EPV was designed for.


\subsection{Spatial Effects $\xi$}
\label{subsec:spat_effects}
Player-tracking data is a breakthrough because it allows us to model the fundamental spatial component of basketball. In our models, we incorporate the properties of court space in spatial effects $\xi_j^{\ell}, \tilde{\xi}_j^{\ell}, \xi_\shot^{\ell}$, which are unknown real-valued functions on $\Ss$, and therefore infinite dimensional. We represent such spatial effects using Gaussian processes (see \citeasnoun{rasmussen2006gaussian} for an overview of modeling aspects of Gaussian processes). Gaussian processes are usually specified by a mean function and covariance function; this approach is computationally intractable for large data sets, as the computation cost of inference and interpolating the surface at unobserved locations is $\mathcal{O}(n^3)$, where $n$ is the number of different points at which $\xi_j^{\ell}$ is observed (for many spatial effects $\xi_j^{\ell}$, the corresponding $n$ would be in the hundreds of thousands). We instead provide $\xi$ with a low-dimensional representation using functional bases \cite{higdon2002space,quinonero2005unifying}, which offers three important advantages. First, this representation is more computationally efficient for large data sets such as ours. Second, functional bases allow for a non-stationary covariance structure that reflects unique spatial dependence patterns on the basketball court. Finally, the finite basis representation allows us to apply the same between-player CAR prior to estimate each player's spatial effects.


Our functional basis representation of a Gaussian process $\xi_j^{\ell}$ relies on $d$ deterministic basis functions $\phi_{j1}, \ldots, \phi_{jd}: \Ss \rightarrow \R$ such that for any $\bz \in \Ss$,
\begin{equation}\label{GP-basis}
\xi_j^{\ell}(\bz) = \sum_{i=1}^d w^{\ell}_{ji}\phi_{ji}(\bz),
\end{equation}
where $\mathbf{w}^{\ell}_j = (w^{\ell}_{j1} \: \ldots \: w^{\ell}_{jd})'$ is a random vector of loadings, $\mathbf{w}^{\ell}_j \sim \N(\boldsymbol{\omega}^{\ell}_j, \boldsymbol{\Sigma}^{\ell}_j)$. Letting $\Phi_j(\bz) = (\phi_{j1}(\bz) \: \ldots \: \phi_{jd}(\bz))'$, we can see that $\xi_j^{\ell}$ given by \eqref{GP-basis} is a Gaussian process with mean function $\Phi_j(\bz)'\boldsymbol{\omega}^{\ell}_j$ and covariance function $\text{Cov}[\xi_j^{\ell}(\bz_1), \xi_j^{\ell}(\bz_2)] =  \Phi_j(\bz_1)'\boldsymbol{\Sigma}^{\ell}_j\Phi_j(\bz_2)$. However, since bases $\phi_{ji}$ are deterministic, each $\xi^{\ell}_j$ is represented as a $d$-dimensional parameter. Note that we also use \eqref{GP-basis} for pass receiver spatial effects and the spatial effect term in the shot probability model, $\tilde{\xi}^{\ell}_j$ and $\xi_\shot^{\ell}$, respectively. For these terms we have associated bases $\tilde{\phi}_{ji}, \phi_{\shot i}$ and weights, $\tilde{w}^{\ell}_{ji}, w^{\ell}_{\shot i}$. As our notation indicates, bases functions $\Phi_j(\bz)$ differ for each macrotransition type but are constant across players; whereas weight vectors $\mathbf{w}^{\ell}_j$ vary across both macrotransition types and players. 

\begin{figure}[t]
\centering
\includegraphics[width=1.0\textwidth]{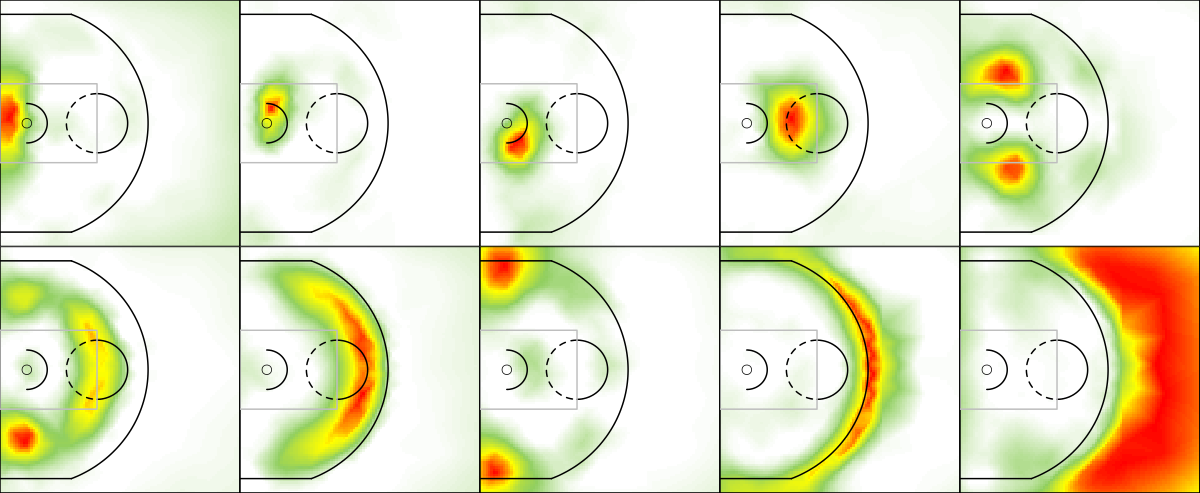}
\caption{The functional bases $\phi_{ji}$ for $i=1, \ldots, 10$ and $j$ corresponding to the shot-taking macrotransition, $j=5$. There is no statistical interpretation of the ordering of the bases; we have displayed them in rough order of the shot types represented, from close-range to long-range.}
\label{bases}
\end{figure}

Using $d=10$, we determine the functional bases in a pre-processing step, discussed in Appendix \ref{subsec:psi}. These basis functions are interpretable as patterns/motifs that constitute players' decision-making tendencies as a function of space; please see Figure \ref{bases}  for an example, or \citeasnoun{miller2013icml} for related work in a basketball context. Furthermore, we use a CAR model \eqref{car} to supply the prior mean and covariance matrix ($\boldsymbol{\omega}^{\ell}_j, \boldsymbol{\Sigma}^{\ell}_j$) for the weights: 
\begin{align}
\boldsymbol{w}^{\ell}_j | \boldsymbol{w}^{-(\ell)}_j, \tau_{\mathbf{w}_j}^2 &\sim 
 \norm \left( \frac{1}{n_{\ell}} \sum_{k : H_{\ell k} = 1} \mathbf{w}^{k}_j, \frac{\tau_{\mathbf{w}_j}^2}{n_{\ell}} \mathbf{I}_d\right) \nonumber \\
\tau^2_{\mathbf{w}_{j}} &\sim \text{InvGam}(1, 1).
\label{car3}
\end{align}
As with \eqref{car},
we also use \eqref{car3} for terms $\tilde{\mathbf{w}}_{j}$ and $\mathbf{w}_\shot$. Combining the functional basis representation \eqref{GP-basis} with the between-player CAR prior \eqref{car3} for the weights, we get a prior representation for spatial effects $\xi^{\ell}_j, \tilde{\xi}^{\ell}_j, \tilde{\xi}^{\ell}$ that is low-dimensional and shares information both across space and between different players.

\subsection{Parameter Estimation}
\label{subsec:estimation}

As discussed in Section \ref{sec:Macro}, calculating EPV requires the parameters that define the multiresolution transition models \ref{M3}--\ref{M4}---specifically, the hazards $\lambda_j^{\ell}$, shot probabilities $p^{\ell}$, and all parameters of the microtransition model \ref{M3}. We estimate these parameters in a Bayesian fashion, combining the likelihood of the observed optical tracking data with the prior structure discussed earlier in this section. Using our multiresolution models, we can write the likelihood for the full optical tracking data, indexed arbitrarily by $t$:
\begin{align} \label{partial}
\prod_{t} \prob(Z_{t + \epsilon} | \Fz_t) & = 
    \Bigg( \overbrace{\prod_{t}  \prob(Z_{t + \epsilon}|M(t)^c, \Fz_t)^{\mathbf{1}[M(t)^c]}}^{L_{\text{mic}}} \Bigg) \Bigg( \overbrace{\prod_{t}\prod_{j=1}^6 \prob(Z_{t + \epsilon}|M_j(t), C_{\delta_t}, \Fz_t)^{\mathbf{1}[M_j(t)]}}^{L_{\text{rem}}}\Bigg) \nonumber \\
& \hspace{-2.5cm} \times \Bigg( \underbrace{\prod_{t} \prob(M(t)^c | \Fz_t)^{\mathbf{1}[M(t)^c]} \prod_{j=1}^6 \prob(M_j(t)|\Fz_t)^{\mathbf{1}[M_j(t)]}}_{L_{\text{entry}}} \Bigg) \Bigg( \underbrace{\prod_t \prod_{j=1}^6 \prob(C_{\delta_t} | M_j(t), \Fz_t)^{\mathbf{1}[M_j(t)]}}_{L_{\text{exit}}} \Bigg),
\end{align}
The factorization used in \eqref{partial} highlights data features that inform different parameter groups: $L_{\text{mic}}$ is the likelihood term corresponding to the microtransition model \ref{M3}, $L_{\text{entry}}$ the macrotransition entry model \ref{M1}, and $L_{\text{exit}}$ the macrotransition exit model \ref{M2}.
The remaining term $L_{\text{rem}}$ is left unspecified, and ignored during inference. 
Thus, $L_{\text{mic}}, L_{\text{entry}},$ and $L_{\text{exit}}$ can be though of as partial likelihoods \cite{cox1975partial}, which under mild conditions leads to consistent and asymptotically well-behaved estimators \cite{wong1986theory}. When parameters in these partial likelihood terms are given prior distributions, as is the case for those comprising the hazards in the macrotransition entry model, as well as those in the shot probability model, the resulting inference is partially Bayesian \cite{cox1975note}.  

The microtransition partial likelihood term $L_{\text{mic}}$ factors by player:
\begin{equation}
L_{\text{mic}} \propto \prod_t \prod_{\ell = 1}^L \prob(\mathbf{z}_{\ell}(t  + \epsilon) | M(t)^c, \Fz_t)^{\mathbf{1}[M(t)^c \text{ and } \ell \text{ on court at time } t]}.
\label{Lmic}
\end{equation}
Depending on whether or not player $\ell$ is on offense (handling the ball or not) or defense, $\prob(\bz_{\ell}(t + \epsilon) | M(t)^c, \Fz_t)$ is supplied by the offensive \eqref{micro} or defensive\eqref{micro-defense} microtransition models. Parameters for these models \eqref{micro}--\eqref{micro-defense} are estimated using R-INLA, where spatial acceleration effects $\mu^{\ell}_x, \mu^{\ell}_y$ are represented using a Gaussian Markov random field approximation to a Gaussian process with Mat\'ern covariance \cite{lindgren2011explicit}. Appendix \ref{subsec:psi} provides the details on this approximation. We do perform any hierarchical modeling for the parameters of the microtransition model---because this model only describes movement (not decision-making), the data for every player is informative enough to provide precise inference. Thus, microtransition models are fit in parallel using each player's data separately; this requires $L=461$ processors, each taking at most 18 hours at 2.50Ghz clock speed, using 32GB of RAM.

For the macrotransition entry term, we can write
\begin{equation}
L_{\text{entry}} \propto \prod_{l=1}^L \prod_{j=1}^6 L_{\text{entry}_j}^{\ell} (\lambda_j^{\ell}(\cdot)),
\label{Lmac}
\end{equation}
recognizing that (for small $\epsilon$),
\begin{align}\label{everything}
L^{\ell}_{\text{entry}_j}(\lambda_j^{\ell}(\cdot))
&\propto
\left(\prod_{\substack{t \: : \: M_j(t) \\ t \in \mathcal{T}^{\ell}}} \lambda^{\ell}_j(t) \right) \exp \left( - \sum_{\substack{
t \in \mathcal{T}^{\ell}}} \lambda^{\ell}_j(t) \right) \nonumber \\
\text{where}\hspace{1cm} \log(\lambda^{\ell}_j(t)) &=
[\mathbf{W}_j^{\ell}(t)]'\boldsymbol{\beta}_j^{\ell} + 
\boldsymbol{\phi}_j(\bz_{\ell}(t))'\mathbf{w}_j^{\ell} + \left(\tilde{\boldsymbol{\phi}}_j(\bz_{j}(t) )' \tilde{\mathbf{w}}_j^{\ell}\mathbf{1}[j \leq 4]\right)
\end{align}
and $\mathcal{T}^{\ell}$ is the set of time $t$ for which player $\ell$ possesses the ball.
Expression \eqref{everything} is the likelihood for a Poisson regression; combined with prior distributions \eqref{car}--\eqref{car3}, inference for $\boldsymbol{\beta}^{\ell}_j, \mathbf{w}_j^{\ell}, \tilde{\mathbf{w}}_j^{\ell}$ is thus given by a hierarchical Poisson regression. However, the size of our data makes implementing such a regression model computationally difficult as the design matrix would have 30.4 million rows and a minimum of $L(p_j + d) \geq 5993$ columns, depending on macrotransition type. We perform this regression through the use of integrated nested Laplace approximations (INLA) \cite{rue2009approximate}. 
Each macrotransition type can be fit separately, and requires approximately 24 hours using a single 2.50GHz processor with 120GB of RAM.

Recalling Section \ref{macro_exit}, the macrotransition exit model \ref{M2} is deterministic for all macrotransitions except shot attempts ($j=5$). Thus, $L_{\text{exit}}$ only provides information on the parameters of our shot probability model \eqref{shotprob}. Analogous to the Poisson model in \eqref{everything}, $L_{\text{exit}}$ is the likelihood of a logistic regression, which factors by player. We also use INLA to fit this hierarchical logistic regression model, though fewer computational resources are required as this likelihood only depends on time points where a shot is attempted, which is a much smaller subset of our data.

%% file: EPV_results.tex
After obtaining parameter estimates for the multresolution transition models, we can calculate EPV using Theorem \ref{epvtheorem} and plot $\nu_t$ throughout the course of any possession in our data. We view such (estimated) EPV curves as the main contribution of our work, and their behavior and potential inferential value has been introduced in Section \ref{sec:Intro}. We illustrate this value by revisiting the possession highlighted in Figure \ref{heat_poss} through the lens of EPV. Analysts may also find meaningful aggregations of EPV curves that summarize players' behavior over a possession, game, or season in terms of EPV. We offer two such aggregations in this section.

\subsection{Predictive Performance of EPV}

Before analyzing EPV estimates, it is essential to check that such estimates are properly calibrated \cite{gneiting2007probabilistic} and accurate enough to be useful to basketball analysts. Our paper introduces EPV, and as such there are no existing results to benchmark the predictive performance of our estimates. We can, however, compare the proposed implementation for estimating EPV with simpler models, based on lower resolution information, to verify whether our multiresolution model captures meaningful features of our data. Assessing the predictive performance of an EPV estimator is difficult because the estimand is a curve whose length varies by possession. Moreover, we never observe any portion of this curve; we only know its endpoint. Therefore, rather than comparing estimated EPV curves between our method and alternative methods, we compare estimated transition probabilities. For any EPV estimator method that is stochastically consistent, if the predicted transitions are properly calibrated, then the derived EPV estimates should be as well.

For the inference procedure in Section \ref{sec:Computation}, we use only 90\% of our data set for parameter inference, with the remaining 10\% used to evaluate the out-of-sample performance of our model. We also evaluated out-of-sample performance of simpler macrotransition entry/exit models, which use varying amounts of information from the data. Table~\ref{loglik-table} provides the out-of-sample log-likelihood for the macrotransition models applied to the 10\% of the data not used in model fitting for various simplified models. In particular, we start with the simple model employing constant hazards for each player/event type, then successively add situational covariates, spatial information, then full hierarchical priors. Without any shrinkage, our full model performs in some cases worse than a model with no spatial effects included, but with shrinkage, it consistently performs the best (highest log-likelihood) of the configurations compared. This behavior justifies the prior structure introduced in Section \ref{sec:Computation}.


\begin{table}[ht]
\centering
\begin{tabular}{lrrrr}
  \toprule
 & \multicolumn{4}{c}{Model Terms} \\
\midrule
Macro. type & Player & Covariates & Covariates + Spatial & Full \\ 
  \midrule
Pass1 & -29.4 & -27.7 & -27.2 & -26.4 \\ 
Pass2 & -24.5 & -23.7 & -23.2 & -22.2 \\ 
Pass3 & -26.3 & -25.2 & -25.3 & -23.9 \\ 
Pass4 & -20.4 & -20.4 & -24.5 & -18.9 \\ 
Shot Attempt & -48.9 & -46.4 & -40.9 & -40.7 \\ 
Made Basket & -6.6 & -6.6 & -5.6 & -5.2 \\ 
Turnover & -9.3 & -9.1 & -9.0 & -8.4 \\ 
   \bottomrule
\end{tabular}
\caption{Out of sample log-likelihood (in thousands) for macrotransition entry/exit models under various model specifications. ``Player'' assumes constant hazards for each player/event type combination. ``Covariates'' augments this model with situational covariates, $\mathbf{W}^{\ell}_j(t)$ as given in \eqref{hazard-equation}. ``Covariates + Spatial'' adds a spatial effect, yielding \eqref{hazard-equation} in its entirety. Lastly, ``Full'' implements this model with the full hierchical model discussed in Section \ref{sec:Computation}.}
\label{loglik-table}
\end{table}


\subsection{Possession Inference from Multiresolution Transitions} 

\begin{figure}[h!]
\centering
\begin{tabular}{ccc}
\subfloat[]{\includegraphics[width=0.325\textwidth]{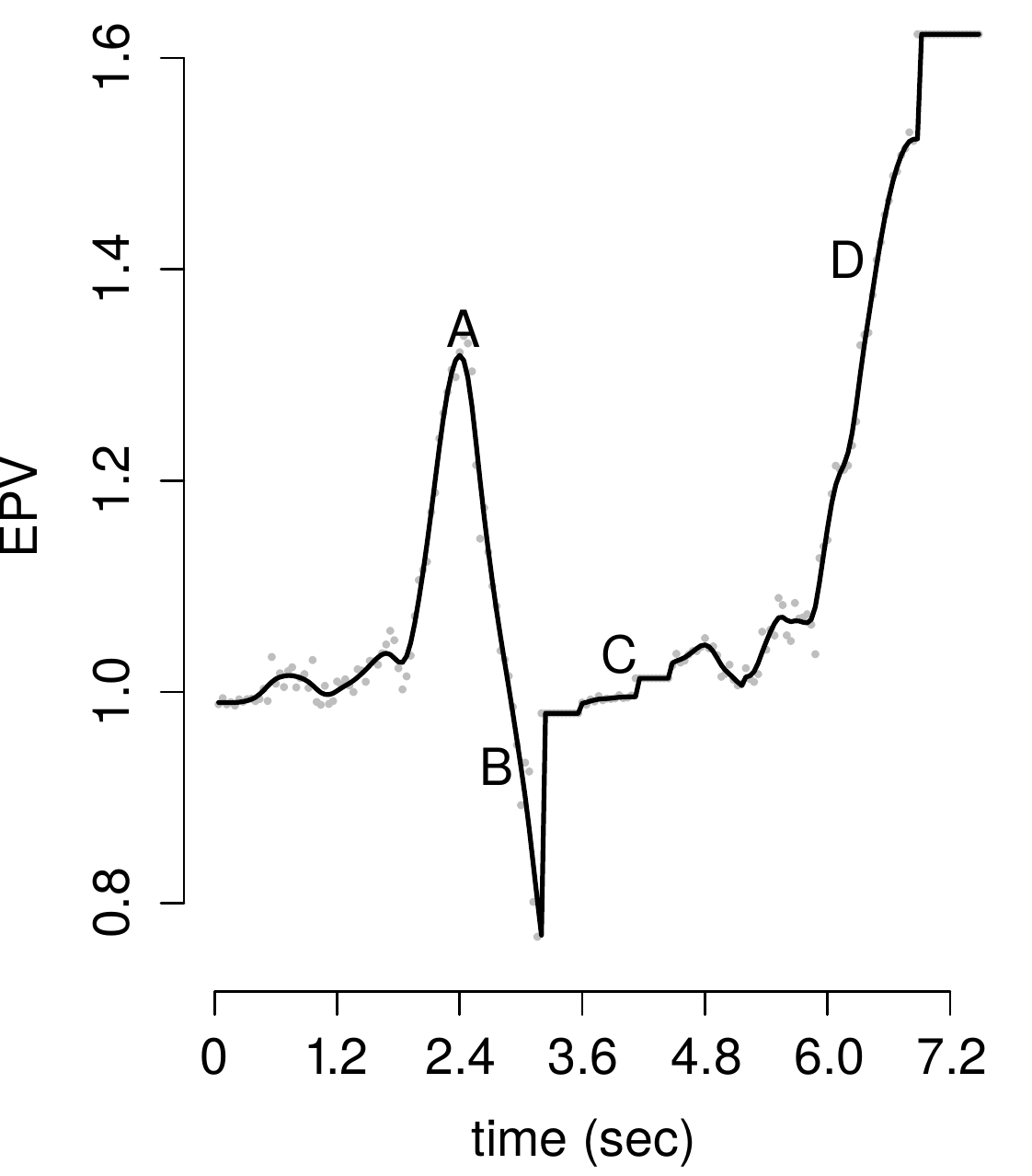}} &
\subfloat[]{\includegraphics[width=0.325\textwidth]{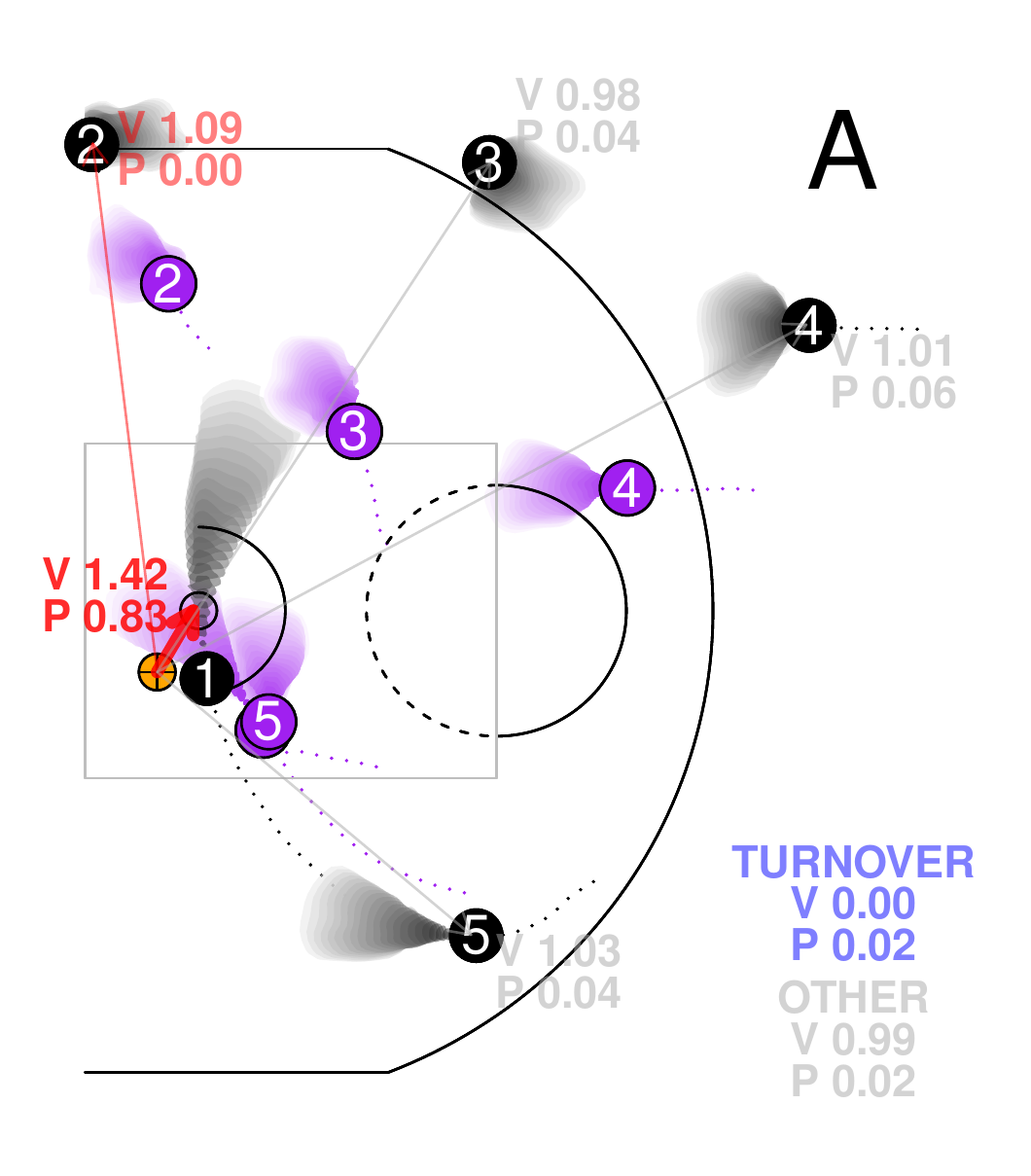}} &
\subfloat[]{\includegraphics[width=0.325\textwidth]{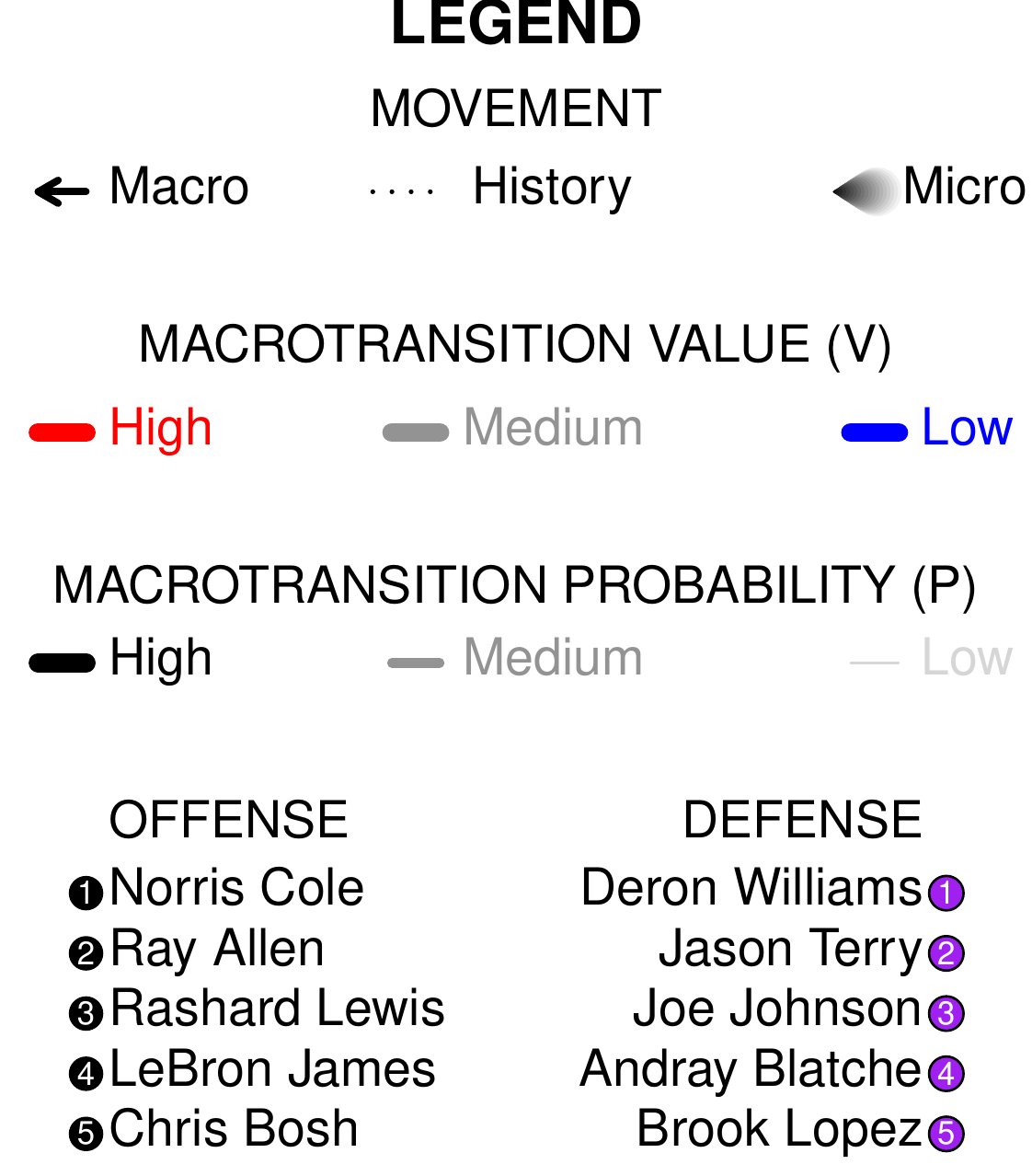}} \\
\subfloat[]{\includegraphics[width=0.325\textwidth]{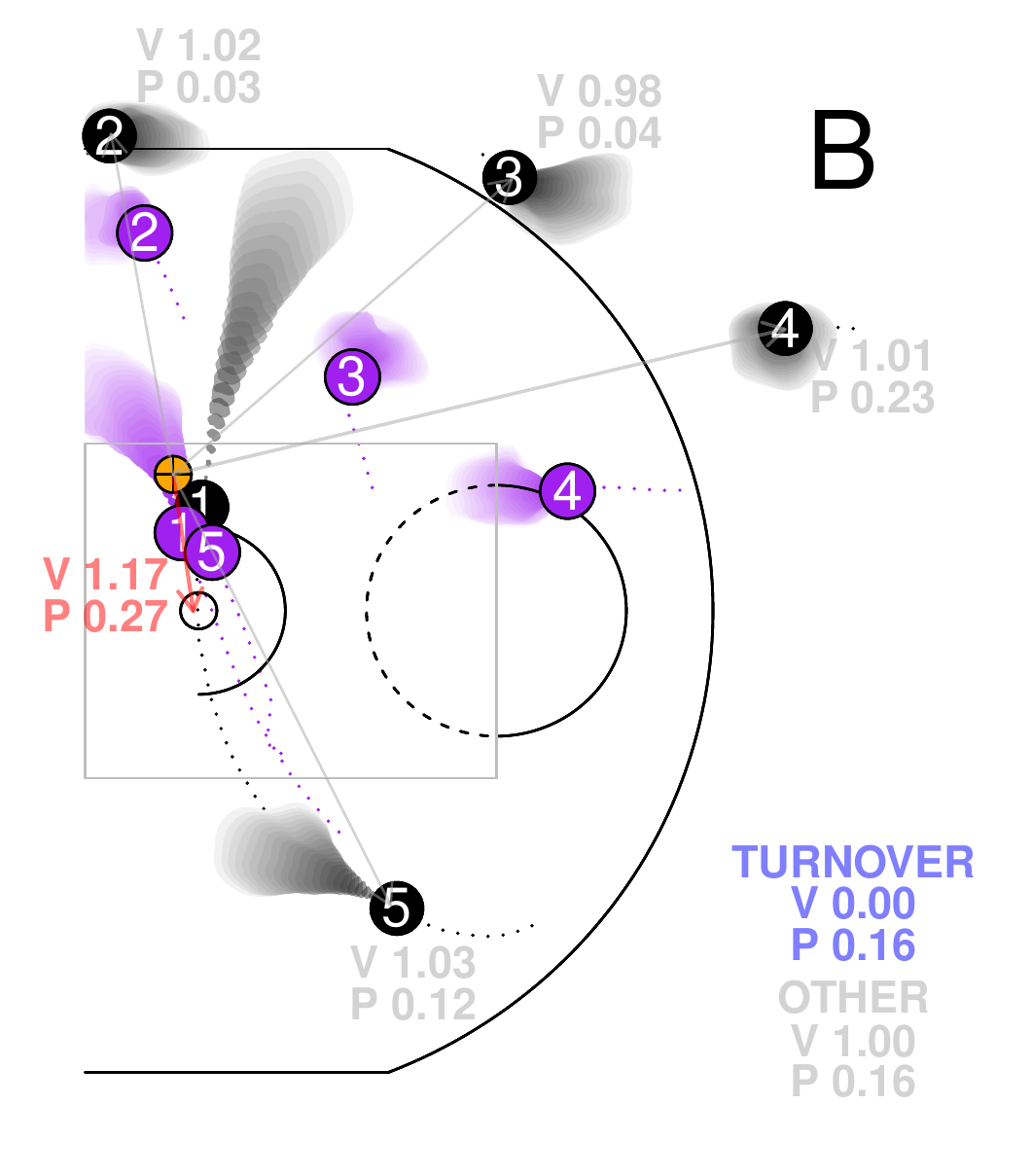}} &
\subfloat[]{\includegraphics[width=0.325\textwidth]{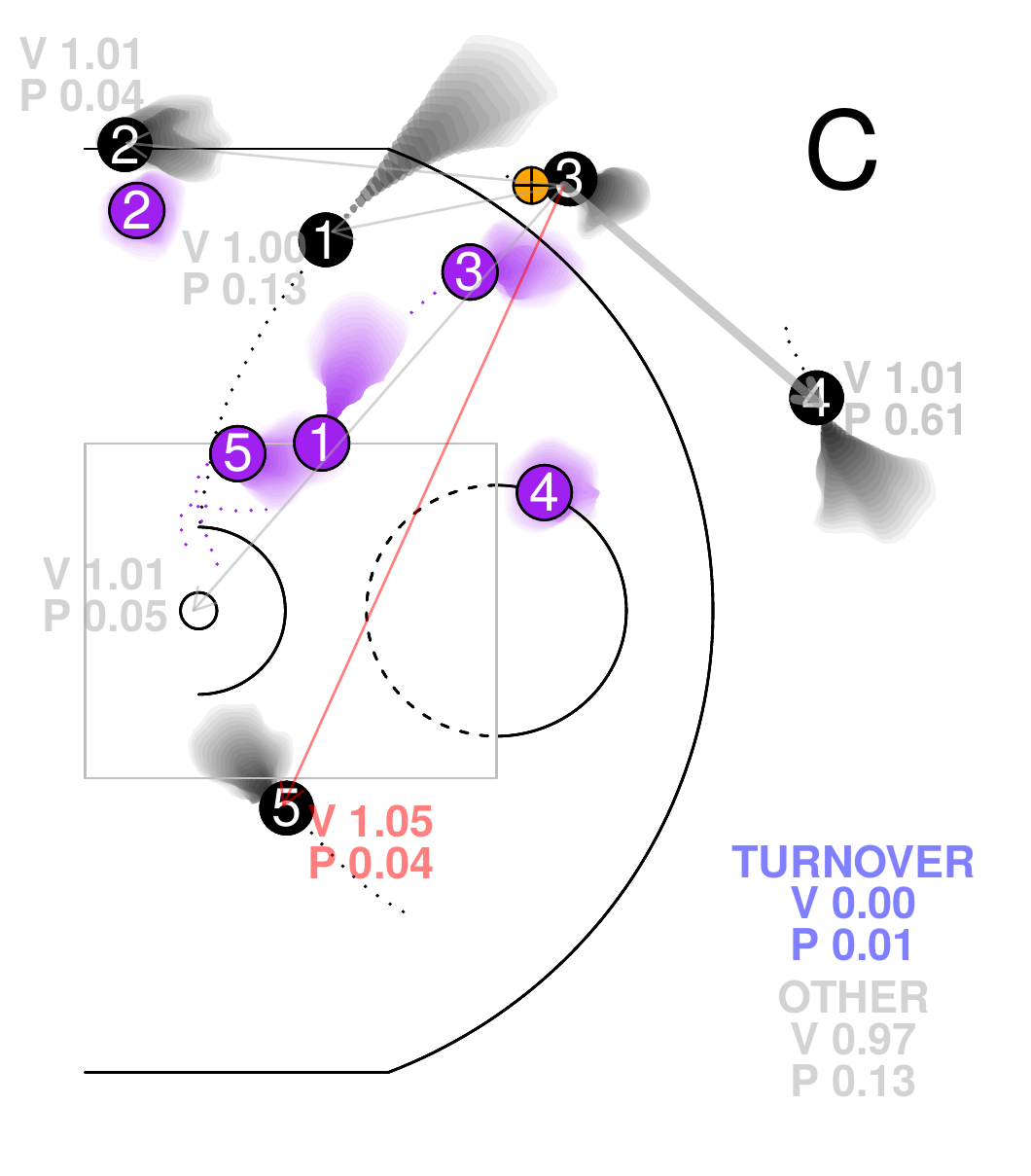}} &
\subfloat[]{\includegraphics[width=0.325\textwidth]{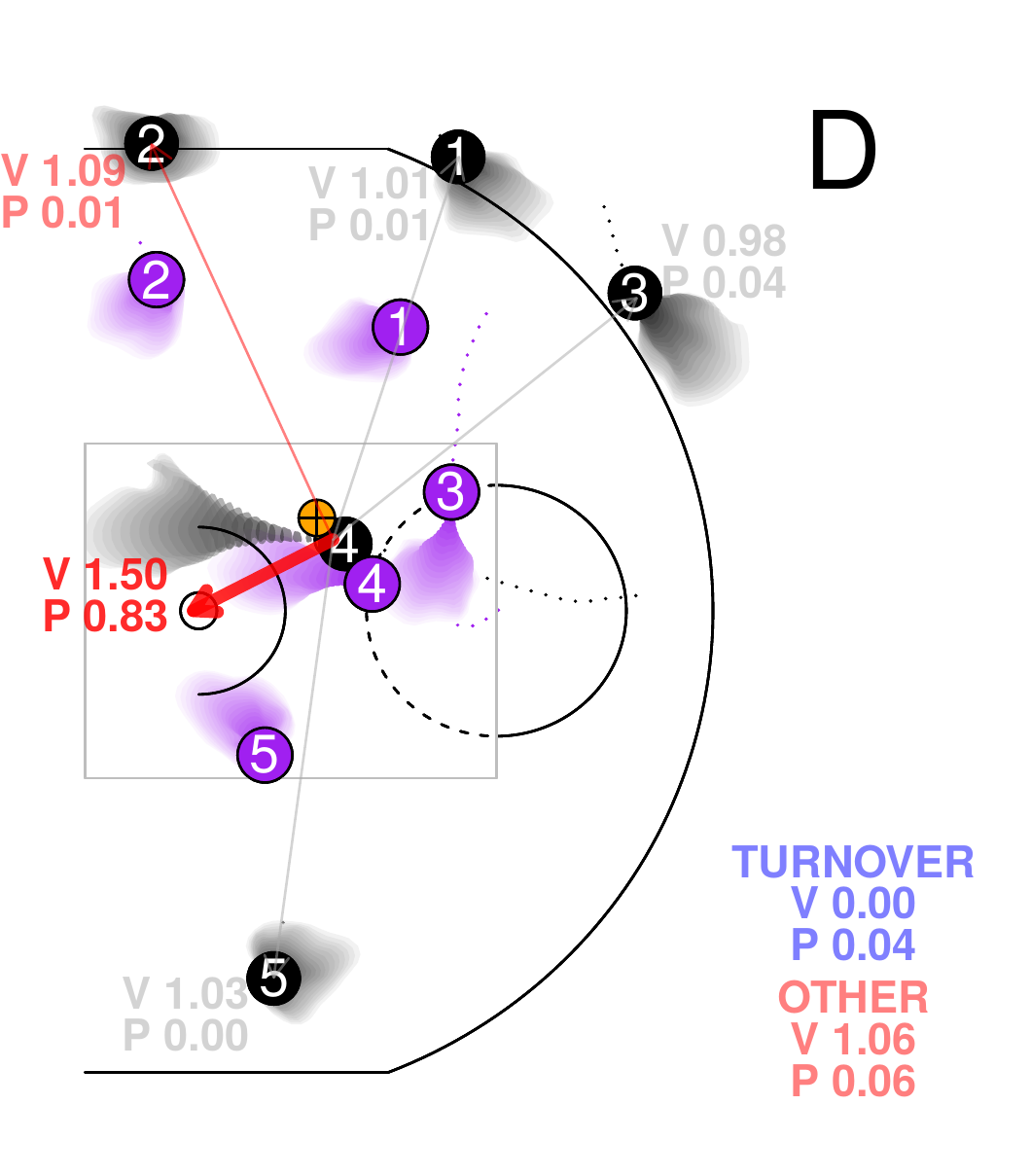}}
\end{tabular}
\caption{Detailed diagram of EPV as a function of multiresolution transition probabilities for four time points (labeled A,B,C,D) of the possession featured in Figures \ref{heat_poss}--\ref{heat_epv}. Two seconds of microtransitions are shaded (with forecasted positions for short time horizons darker) while macrotransitions are represented by arrows, using color and line thickness to encode the value (V) and probability (P) of such macrotransitions. The value and probability of the ``other'' category represents the case that no macrotransition occurs during the next two seconds.}
\label{heat_detail}
\end{figure}

Understanding the calculation of EPV in terms of multiresolution transitions is a valuable exercise for a basketball analyst, as these model components reveal precisely how the EPV estimate derives from the spatiotemporal circumstances of the time point considered. Figure \ref{heat_detail} diagrams four moments during our example possession (introduced originally in Figures \ref{heat_poss} and \ref{heat_epv}) in terms of multiresolution transition probabilities. These diagrams illustrate Theorem \ref{epvtheorem} by showing EPV as a weighted average of the value of the next macrotransition. Potential ball movements representing macrotransitions are shown as arrows, with their respective values and probabilities graphically illustrated by color and line thickness (this information is also annotated explicitly). Microtransition distributions are also shown, indicating distributions of players' movement over the next two seconds. Note that the possession diagrammed here was omitted from the data used for parameter estimation.
\captionsetup[subfigure]{labelformat=empty}


Analyzing Figure \ref{heat_detail}, we see that our model estimates largely agree with basketball intuition. For example, players are quite likely to take a shot when they are near to and/or moving towards the basket, as shown in panels A and D. Additionally, because LeBron James is a better shooter than Norris Cole, the value of his shot attempt is higher, even though in the snapshot in panel D he is much farther from the basket than Cole is in panel A. While the value of the shot attempt averages over future microtransitions, which may move the player closer to the basket, when macrotransition hazards are high this average is dominated by microtransitions on very short time scales.

We also see Ray Allen, in the right corner 3, as consistently one of the most valuable pass options during this possession, particularly when he is being  less closely defended as in panels A and D. In these panels, though, we never see an estimated probability of him receiving a pass above 0.05, most likely because he is being fairly closely defended for someone so far from the ball, and because there are always closer passing options for the ballcarrier. Similarly, while Chris Bosh does not move much during this possession, he is most valuable as a passing option in panel C where he is closest to the basket and without any defenders in his lane. 
From this, we see that the estimated probabilities and values of the macrotransitions highlighted in Figure \ref{heat_detail} match well with basketball intuition.

The analysis presented here could be repeated on any of hundreds of thousands of possessions available in a season of optical tracking data. EPV plots as in Figure \ref{heat_epv} and diagrams as in Figure \ref{heat_detail} provide powerful insight as to how players' movements and decisions contribute value to their team's offense. With this insight, coaches and analysts can formulate strategies and offensive schemes that make optimal use of their players' ability---or, defensive strategies that best suppress the motifs and situations that generate value for the opposing offense.

\subsection{EPV-Added}

Aggregations of EPV estimates across possessions can yield useful summaries for player evaluation. For example, \textit{EPV-Added} (EPVA) quantifies a player's overall offensive value through his movements and decisions while handling the ball, relative to the estimated value contributed by a league-average player receiving ball possession in the same situations. The notion of \textit{relative} value is important because the martingale structure of EPV ($\nu_t$) prevents any meaningful aggregation of EPV across a specific player's possessions. $\E[\nu_t - \nu_{t + \epsilon}] = 0$ for all $t$, meaning that \textit{on average} EPV does not change during any specific player's ball handling. Thus, while we see the EPV skyrocket after LeBron James receives the ball and eventually attack the basket in Figure \ref{heat_epv}, the definition of EPV prevents such increases being observed on average. 

If player $\ell$ has possession of the ball starting at time $t_s$ and ending at $t_e$, the quantity $\nu_{t_e} - \nu_{t_s}^{r(\ell)}$ estimates the value contributed player by $\ell$ relative to the hypothetical league-average player during his ball possession (represented by $\nu_{t_s}^{r(\ell)}$). We calculate EPVA for player $\ell$ (EPVA($\ell$)) by summing such differences over all a player's touches (and dividing by the number of games played by player $\ell$ to provide standardization):
\begin{equation}\label{EPVA}
\text{EPVA}(\ell) = \frac{1}{\# \text{ games for $\ell$}}\sum_{\{t_s, t_e\} \in \mathcal{T}^{\ell}} \nu_{t_e} - \nu_{t_s}^{r(\ell)}
\end{equation}
where $\mathcal{T}^{\ell}$ contains all intervals of form $[t_s, t_e]$ that span player $\ell$'s ball possession. Specific details on calculating $\nu_t^{r(\ell)}$ are included in Appendix \ref{subsec:EPVA}. 

Averaging over games implicitly rewards players who have high usage, even if their value added per touch might be low. Often, one-dimensional offensive players accrue the most EPVA per touch since they only handle the ball when they are uniquely suited to scoring; for instance, some centers (such as the Clippers' DeAndre Jordan) only receive the ball right next to the basket, where their height offers a considerable advantage for scoring over other players in the league. Thus, averaging by game---not touch---balances players' efficiency per touch with their usage and importance in the offense. Depending on the context of the analysis, EPVA can also be adjusted to account for team pace (by normalizing by 100 possession) or individual usage (by normalizing by player-touches).

Table~\ref{epv_tab} provides a list of the top and bottom 10 ranked players by EPVA using our 2013-14 data. Generally, players with high EPVA effectively adapt their decision-making process to the spatiotemporal circumstances they inherit when gaining possession. They receive the ball in situations that are uniquely suited to their abilities, so that on average the rest of the league is less successful in these circumstances. Players with lower EPVA are not necessarily ``bad'' players in any conventional sense; their actions simply tend to lead to fewer points than other players given the same options. Of course, EPVA provides a limited view of a player's overall contributions since it does not quantify players' actions on defense, or other ways that a player may impact EPV while not possessing the ball (though EPVA could be extended to include these aspects).

\begin{table}[ht]
\centering
\begin{tabular}{rlr}
  \toprule
Rank &  Player & EPVA \\
   \midrule
1 & Kevin Durant & 3.26 \\
2 & LeBron James & 2.96 \\
3 & Jose Calderon & 2.79 \\
4 & Dirk Nowitzki & 2.69 \\
5 & Stephen Curry & 2.50 \\
6 & Kyle Korver & 2.01 \\
7 & Serge Ibaka & 1.70 \\
8 & Channing Frye & 1.65 \\
9 & Al Horford & 1.55 \\
10 & Goran Dragic & 1.54 \\
\bottomrule
\end{tabular}
\quad
\begin{tabular}{rlr}
  \toprule
Rank & Player & EPVA \\ 
  \midrule
277 & Zaza Pachulia & -1.55 \\
278 & DeMarcus Cousins & -1.59 \\
279 & Gordon Hayward & -1.61 \\
280 & Jimmy Butler & -1.61 \\
281 & Rodney Stuckey & -1.63 \\
282 & Ersan Ilyasova & -1.89 \\
283 & DeMar DeRozan & -2.03 \\
284 & Rajon Rondo & -2.27 \\
285 & Ricky Rubio & -2.36 \\
286 & Rudy Gay & -2.59 \\
\bottomrule
\end{tabular}
\caption[]{Top/bottom 10 players by EPVA per game in 2013-14, minimum 500 touches in season.}
\label{epv_tab}
\end{table}

As such, we stress the idea that EPVA is not a best/worst players in the NBA ranking. Analysts should also be aware that the league-average player being used
as a baseline is completely hypothetical, and we heavily extrapolate our model output by
considering value calculations assuming this nonexistant player possessing the ball in all the
situations encountered by an actual NBA player. The extent to which such an extrapolation is valid is a judgment a basketball expert can make. Alternatively, one can consider EPV-added over \textit{specific} players (assuming player $\ell_2$ receives the ball in the same situations as player $\ell_1$), using the same framework developed for EPVA. Such a quantity may actually be more useful, particularly if the players being compared play similar roles on their teams and face similar situations and the degree of extrapolation is minimized. 

\subsection{Shot Satisfaction}

Aggregations of the individual components of our multiresolution transition models can also provide useful insights. For example, another player metric we consider is called \textit{shot satisfaction}. For each shot attempt a player takes, we wonder how satisfied the player is with his decision to shoot---what was the expected point value of his most reasonable passing option at the time of the shot? If for a particular player, the EPV measured at his shot attempts is higher than the EPV conditioned on his possible passes at the same time points, then by shooting the player is usually making the best decision for his team. On the other hand, players with pass options at least as valuable as shots should regret their shot attempts (we term ``satisfaction'' as the opposite of regret) as passes in these situations have higher expected value. 

\begin{table}[h!]
\centering
\begin{tabular}{rlr}
  \toprule
Rank &  Player & Shot Satis. \\
   \midrule
1 & Mason Plumlee & 0.35 \\
2 & Pablo Prigioni & 0.31 \\
3 & Mike Miller & 0.27 \\
4 & Andre Drummond & 0.26 \\
5 & Brandan Wright & 0.24 \\
6 & DeAndre Jordan & 0.24 \\
7 & Kyle Korver & 0.24 \\
8 & Jose Calderon & 0.22 \\
9 & Jodie Meeks & 0.22 \\
10 & Anthony Tolliver & 0.22 \\
\bottomrule
\end{tabular}
\quad
\begin{tabular}{rlr}
  \toprule
Rank & Player & Shot Satis. \\ 
  \midrule
277 & Garrett Temple & -0.02 \\
278 & Kevin Garnett & -0.02 \\
279 & Shane Larkin & -0.02 \\
280 & Tayshaun Prince & -0.03 \\
281 & Dennis Schroder & -0.04 \\
282 & LaMarcus Aldridge & -0.04 \\
283 & Ricky Rubio & -0.04 \\
284 & Roy Hibbert & -0.05 \\
285 & Will Bynum & -0.05 \\
286 & Darrell Arthur & -0.05 \\
\bottomrule
\end{tabular}
\caption[]{Top/bottom 10 players by shot satisfaction in 2013-14, minimum 500 touches in season.}
\label{satis_tab}
\end{table}

Specifically, we calculate 
\begin{equation}\label{satisfaction}
\text{SATIS}(\ell) = \frac{1}{|\mathcal{T}^{\ell}_{\text{shot}}|} \sum_{t \in \mathcal{T}^{\ell}_{\text{shot}}} \nu_t - \E\left[X \mid \bigcup_{j=1}^4 M_j(t), \Fz_t \right]
\end{equation}
where $\mathcal{T}^{\ell}_{\text{shot}}$ indexes times a shot attempt occurs, $\{t : M_5(t) \}$, for player $\ell$. Recalling that macrotransitions $j=1, \ldots, 4$ correspond to pass events (and $j=5$ a shot attempt), $\bigcup_{j=1}^4 M_j(t)$ is equivalent to a pass happening in $(t, t + \epsilon]$. Unlike EPVA, shot satisfaction SATIS($\ell$) is expressed as an average per shot (not per game), which favors player such as three point specialists, who often take fewer shots than their teammates, but do so in situations where their shot attempt is extremely valuable. Table \ref{satis_tab} provides the top/bottom 10 players in shot satisfaction for our 2013-14 data. While players who mainly attempt three-pointers (e.g. Miller, Korver) and/or shots near the basket (e.g. Plumlee, Jordan)  have the most shot satisfaction, players who primarily take mid-range or long-range two-pointers (e.g. Aldridge, Garnett) or poor shooters (e.g. Rubio, Prince) have the least. However, because shot satisfaction numbers are mostly positive league-wide, players still shoot relatively efficiently---almost every player generally helps his team by shooting rather than passing in the same situations, though some players do so more than others.

We stress that the two derived metrics given in this paper, EPVA and shot satisfaction, are simply examples of the kinds of analyses enabled by EPV. Convential metrics currently used in basketball analysis do measure shot selection and efficiency, as well as passing rates and assists, yet EPVA and shot satisfaction are novel in analyzing these events in their spatiotemporal contexts. 

%% file: EPV_discuss.tex
This paper introduces a new quantity, EPV, which represents a paradigm shift in the possibilities for statistical inferences about basketball. Using high resolution, optical tracking data, EPV reveals the value in many of the schemes and motifs that characterize basketball offenses but are omitted in the box score. For instance, as diagrammed in Figures \ref{heat_epv} and \ref{heat_detail}, we see that EPV may rise as a player attacks the basket (more so for a strong scorer like LeBron James than for a bench player like Norris Cole), passes to a well-positioned teammate, or gains separation from the defense. Aside from simply tracking changes in EPV, analysts can understand why EPV changes by expressing its value as a weighted average of transition values (as done in Figure~\ref{heat_detail}). Doing so reveals that the source of a high (or low) EPV estimate may come from alternate paths of the possession that were never realized, but were probable enough to have influenced the EPV estimate---an open teammate in a good shooting location, for instance. These insights, which can be reproduced for any valid NBA possession in our data set, have the potential to reshape the way we quantify players' actions and decisions.

We make a number of assumptions---mostly to streamline and simplify our modeling and analysis pipeline---that could be relaxed and yield a more precise model. The largest assumption is that the particular coarsened view of a basketball possession that we propose here is marginally semi-Markov. While this serves as a workable first-order approximation, there are cases that clearly violate this assumption, for example, pre-set plays that string together sequences of runs and passes. Future refinements of the model could define a wider set of macrotransitions and coarsened states to encapsulate these motifs, effectively encoding this additional possession structure from the coach's playbook. 

A number of smaller details could also be addressed. For instance, it seems desirable to model rebound outcomes conditional on high resolution information, such as the identities and motion dynamics of potential rebounders; we do not do this, however, and use a constant probability for each team of a rebound going to either the offense or defense. We also do not distinguish between different types of turnovers (steals, bad passes, ball out of bounds, etc.), though this is due to a technical feature of our data set. Indeed, regardless of the complexity and refinement of an EPV model, we stress that the full resolution data still omits key information, such as the positioning of players' hands and feet, their heights when jumping, and other variables that impact basketball outcomes. As such, analyses based on EPV are best accompanied by actual game film and the insight of a basketball expert.

The computational requirements of estimating EPV curves (and the parameters that generate them) likely limit EPV discussions to academic circles and professional basketball teams with access to the appropriate resources. Our model nevertheless offers a case study whose influence extends beyond basketball. High resolution spatiotemporal data sets are an emerging inferential topic in a number of scientific or business areas, such as climate, security and surveillance, advertising, and gesture recognition. Many of the core methodological approaches in our work, such as using multiresolution transitions and hierarchical spatial models, provide insight beyond the scope of basketball to other spatiotemporal domains.

%% file: EPV_fullspec.tex
In this appendix we provide additional details on steps used in fitting multiresolution models and deriving basketball metrics from EPV estimates. 


\subsection{Time-Varying Covariates in Macrotransition Entry Model}
\label{Covariates}

As revealed in \eqref{hazard-equation}, the hazards $\lambda_j^{\ell}(t)$ are parameterized by spatial effects ($\xi_j^{\ell}$ and $\tilde{\xi}_j^{\ell}$ for pass events), as well as coefficients for situation covariates, $\boldsymbol{\beta}_j^{\ell}$. The covariates used may be different for each macrotransition $j$, but we assume for each macrotransition type the same covariates are used across players $\ell$. 

Among the covariates we consider, \texttt{dribble} is an indicator of whether the ballcarrier has started dribbling after receiving possession. \texttt{ndef} is the distance between the ballcarrier and his nearest defender (transformed to $\log(1 + d)$). \texttt{ball\_lastsec} records the distance traveled by the ball in the previous one second. \texttt{closeness} is a categorical variable giving the rank of the ballcarrier's teammates' distance to the ballcarrier. Lastly, \texttt{open} is a measure of how open a potential pass receiver is using a simple formula relating the positions of the defensive players to the vector connecting the ballcarrier with the potential pass recipient.

For $j \leq 4$, the pass event macrotransitions, we use \texttt{dribble}, \texttt{ndef}, \texttt{closeness}, and \texttt{open}. For shot-taking and turnover events, \texttt{dribble}, \texttt{ndef}, and \texttt{ball\_lastsec} are included. Lastly, the shot probability model (which, from  \eqref{shotprob} has the same parameterization as the macrotransition model) uses \texttt{dribble} and \texttt{ndef} only. All models also include an intercept term. As discussed in Section \ref{subsec:CAR}, independent CAR priors are assumed for each coefficient in each macrotransition hazard model.

\subsection{Player Similarity Matrix $\mathbf{H}$ for CAR Prior}
\label{subsec:H}

The hierarchical models used for parameters of the macrotransition entry model, discussed in Section \ref{subsec:CAR}, are based on the idea that players who share similar roles for their respective teams should behave similarly in the situations they face. Indeed, players' positions (point guard, power forward, etc.) encode their offensive responsibilities: point guards move and distribute the ball, small forwards penetrate and attack the basket, and shooting guards get open for three-point shots. Such responsibilities reflect spatiotemporal decision-making tendencies, and therefore informative for our macrotransition entry model \eqref{hazard-def}--\eqref{hazard-equation}.

\begin{figure}[h]
\centering
\includegraphics[width=1.0\linewidth]{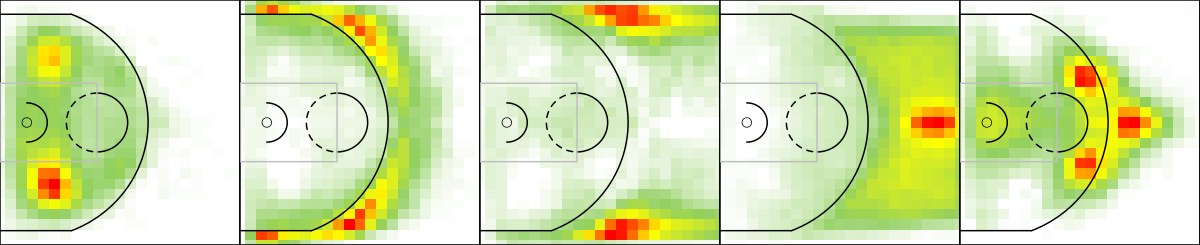}
\caption{The rows of $\mathbf{V}$ (plotted above for $r=5$) are bases for the players' court occupancy distribution. There is no interpretation to the ordering.}
\label{H_bases}
\end{figure}

Rather than use the labeled positions in our data, we define position as a distribution of a player's location during his time on the court. Specifically, we divide the offensive half of the court into 4-square-foot bins (575 total) and count, for each player, the number of data points for which he appears in each bin. Then we stack these counts together into a $L \times 575$ matrix (there are $L=461$ players in our data), denoted $\mathbf{G}$, and take the square root of all entries in $\mathbf{G}$ for normalization. We then perform non-negative matrix factorization (NMF) on $\mathbf{G}$ in order to obtain a low-dimensional representation of players' court occupancy that still reflects variation across players \cite{miller2013icml}. Specifically, this involves solving:
\begin{equation}\label{nmf}
\hat{\mathbf{G}} = \underset{\mathbf{G}^*}{\text{argmin}}\{D(\mathbf{G}, \mathbf{G}^*)\}, \text{ subject to } \mathbf{G}^* = \left(\underset{L \times r}{\mathbf{U}}\right)\left(\underset{r \times 575}{\mathbf{V}}\right) \text{ and } U_{ij},V_{ij} \geq 0 \text{ for all } i,j,
\end{equation}
where $r$ is the rank of the approximation $\hat{\mathbf{G}}$ to $\mathbf{G}$ (we use $r=5$), and $D$ is some distance function; we use a Kullback-Liebler type
$$D(\mathbf{G}, \mathbf{G}^*) = \sum_{i,j} G_{ij}\log \left( G_{ij}/G_{ij}^*\right) - G_{ij} + G_{ij}^*.$$ The rows of $\mathbf{V}$ are non-negative basis vectors for players' court occupancy distributions (plotted in Figure \ref{H_bases}) and the rows of $\mathbf{U}$ give the loadings for each player. With this factorization, $\mathbf{U}_i$ (the $i$th row of $\mathbf{U}$) provides player $i$'s ``position''---a $r$-dimensional summary of where he spends his time on the court. Moreover, the smaller the difference between two players' positions, $||\mathbf{U}_i - \mathbf{U}_j||$, the more alike are their roles on their respective teams, and the more similar we expect the parameters of their macrotransition models to be a priori.

Formalizing this, we take the $L \times L$ matrix $\mathbf{H}$ to consist of 0s, then set $H_{ij} = 1$ if player $j$ is one of the eight closest players in our data to player $i$ using the distance $||\mathbf{U}_i - \mathbf{U}_j||$ (the cutoff of choosing the closest eight players is arbitrary). This construction of $\mathbf{H}$ does not guarantee symmetry, which is required for the CAR prior we use, thus we set $H_{ji} = 1$ if $H_{ij} = 1$. For instance, LeBron James' ``neighbors'' are (in no order): Andre Iguodala, Harrison Barnes, Paul George, Kobe Bryant, Evan Turner, Carmelo Anthony, Rodney Stuckey, Will Barton, and Rudy Gay.


\subsection{Basis Functions for Spatial Effects $\xi$}
\label{subsec:psi}

Recalling \eqref{GP-basis}, for each player $\ell$ and macrotransition type $j$, we have $\xi_j^{\ell}(\bz) = \sum_{i=1}^d w^{\ell}_{ji} \phi_{ji}(\bz)$, where $\{\phi_{ji}, i=1, \ldots, d\}$ are the basis functions for macrotransition $j$. During the inference discussed in Section \ref{sec:Computation}, these basis functions are assumed known. They are derived from a pre-processing step. Heuristically, they are constructed by approximately fitting a simplified macrotransition entry model with stationary spatial effect for each player, then performing NMF to find a low-dimensional subspace (in this function space of spatial effects) that accurately captures the spatial dependence of players' macrotransition behavior. We now describe this process in greater detail.

\begin{figure}[h]
\centering
\includegraphics[width=0.33\linewidth]{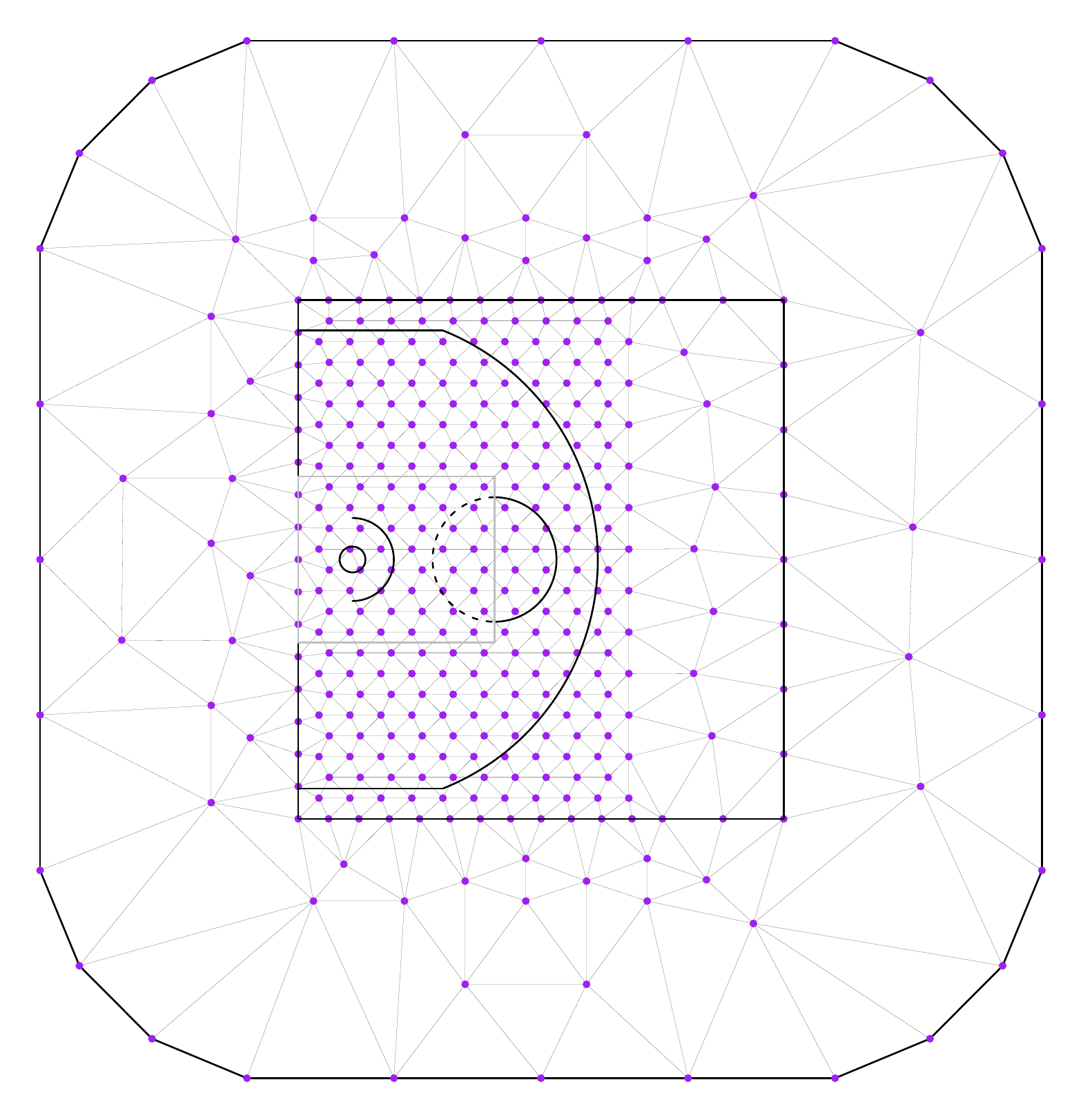}
\caption{Triangulation of $\Ss$ used to build the functional basis $\{\psi_k, k=1, \ldots, d_0\}$. Here, $d_0=383$.}
\label{triangulation}
\end{figure}

Each basis function $\phi_{ji}$ is itself represented as a linear combination of basis functions,
\begin{equation}
\label{phi_basis}
\phi_{ji}(\bz) = \sum_{k=1}^{d_0} v_{jik} \psi_k(\bz),
\end{equation}
where $\{\psi_k, k=1, \ldots, d_0\}$ are basis functions (as the notation suggests, the same basis is used for all $j$, $i$). The basis functions $\{\psi_k, k=1, \ldots, d_0\}$ are induced by a triangular mesh of $d_0$ vertices (we use $d_0 = 383$) on the court space $\Ss$. In practice, the triangulation is defined on a larger region that includes $\Ss$, due to boundary effects. The mesh is formed by partitioning $\Ss$ into triangles, where any two triangles share at most one edge or corner; see Figure \ref{triangulation} for an illustration. With some arbitrary ordering of the vertices of this mesh, $\psi_k:\Ss \rightarrow \R$ is the unique function taking value 0 at all vertices $\tilde{k} \neq k$, 1 at vertex $k$, and linearly interpolating between any two points within the same triangle used in the mesh construction. Thus, with this basis, $\phi_{ji}$ (and consequently, $\xi^{\ell}_j$) are piecewise linear within the triangles of the mesh. 

This functional basis $\{\psi_k, k=1, \ldots, d_0\}$ is used by \citeasnoun{lindgren2011explicit}, who show that it can approximate a Gaussian random field with Mat\'ern covariance. Specifically, let $x(\bz) = \sum_{k=1}^{d_0} v_k\psi_k(\bz)$ and assume $(v_1 \: \ldots \: v_k)' = \mathbf{v} \sim \N(0, \boldsymbol{\Sigma}_{\nu, \kappa, \sigma^2})$. The form of $\boldsymbol{\Sigma}_{\nu, \kappa, \sigma^2}$ is such that the covariance function of $x$ approximates a Mat\'ern covariance:
\begin{equation}
\label{gmrf}
\text{Cov}[x(\bz_1), x(\bz_2)] = \boldsymbol{\psi}(\bz_1)'\boldsymbol{\Sigma}_{\nu, \kappa, \sigma^2} \boldsymbol{\psi}(\bz_2) \approx \frac{\sigma^2}{\Gamma(\nu)2^{\nu-1}}(\kappa ||\bz_1 - \bz_2||)^{\nu} K_{\nu}(\kappa ||\bz_1 - \bz_2||),
\end{equation}
where $\boldsymbol{\psi}(\bz) = (\psi_1(\bz) \: \ldots \: \psi_{d_0}(\bz))'$. As discussed in Section \ref{subsec:spat_effects}, the functional basis representation of a Gaussian process offers computational advantages in that the infinite dimensional field $x$ is given a $d_0$-dimensional representation, as $x$ is completely determined by $\mathbf{v}$. Furthermore, as discussed in \citeasnoun{lindgren2011explicit}, $\boldsymbol{\Sigma}_{\nu, \kappa, \sigma^2}^{-1}$ is sparse (\eqref{gmrf} is actually a Gaussian Markov random field (GMRF) approximation to $x$), offering additional computational savings \cite{rue2001fast}.

The GMRF approximation given by \eqref{phi_basis}--\eqref{gmrf} is actually used in fitting the microtransition models for offensive players \eqref{micro}. We give the spatial innovation terms $\mu^{\ell}_x, \mu^{\ell}_y$ representations using the $\psi$ basis. Then, as mentioned in Section \ref{subsec:estimation}, \eqref{micro} is fit independently for each player in our data set using the software R-INLA.

We also fit simplified versions of the macrotransition entry model, using the $\psi$ basis, in order to determine $\{v_{jik}, k=1, \ldots, d_0\}$, the loadings of the basis representation for $\phi$, \eqref{phi_basis}. This simplified model replaces the macrotransition hazards \eqref{hazard-equation} with
\begin{equation}
\label{hazard_preprocess}
\log(\lambda^{\ell}_j(t)) = c_j^{\ell} + \sum_{k=1}^{d_0} u^{\ell}_{jk} \psi_k(\bz^{\ell}(t))
+ \mathbf{1}[j \leq 4]\sum_{k=1}^{d_0}\tilde{u}_{jk}^{\ell}\psi_k\left(\bz_{j}(t)\right),
\end{equation}
thus omitting situational covariates ($\boldsymbol{\beta}^{\ell}_j$ in \eqref{hazard-equation}) and using the $\psi$ basis representation in place of $\xi_j^{\ell}$. Note that for pass events, like \eqref{hazard-equation}, we have an additional term based on the pass recipient's location, parameterized by $\{\tilde{u}^{\ell}_{jk}, k=1, \ldots, d_0\}$. As discussed in Section \ref{subsec:estimation}, parameters in \eqref{hazard_preprocess} can be estimated by running a Poisson regression. We perform this independently for all players $\ell$ and macrotransition types $j$ using the R-INLA software. Like the microtransition model, we fit \eqref{hazard_preprocess} separately for each player across $L = 461$ processors (each hazard type $j$ is run in serial), each requiring at most 32GB RAM and taking no more than 16 hours.

For each macrotransition type $j$, point estimates $\hat{u}^{\ell}_{jk}$ are exponentiated\footnote{The reason for exponentiation is because estimates $\hat{u}^{\ell}_{jk}$ inform the log hazard, so exponentiation converts these estimates to a more natural scale of interest. Strong negative signals among the $\hat{u}^{\ell}_{jk}$ will move to 0 in the entries of $\mathbf{U}_j$ and not be very influential in the matrix factorization \eqref{U_factorization}, which is desirable for our purposes.}, so that $[\mathbf{U}_j]_{\ell k} = \exp(\hat{u}^{\ell}_{jk})$. We then perform NMF \eqref{nmf} on $\mathbf{U}_j$:
\begin{equation}
\label{U_factorization}
\mathbf{U}_j \approx \left(\underset{L \times d}{\mathbf{Q}_j}\right)\left(\underset{d \times d_0}{\mathbf{V}_j}\right).
\end{equation}
Following the NMF example in Section \ref{subsec:H}, the rows of $\mathbf{V}_j$ are bases for the variation in coefficients $\{u^{\ell}_{jk}, k=1, \ldots, d_0\}$ across players $\ell$. As $1 \leq k \leq d_0$ indexes points on our court triangulation (Figure \ref{triangulation}), such bases reflect structured variation across space. We furthermore use these terms as the coefficients for \eqref{phi_basis}, the functional basis representation of $\phi_{ji}$, setting $v_{jik}  = [\mathbf{V}_j]_{ik}$. Equivalently, we can summarize our spatial basis model as:
\begin{equation}
\label{all_bases}
\xi^{\ell}_j (\bz) = [\mathbf{w}^{\ell}_j]'\boldsymbol{\phi}_j(\bz) = [\mathbf{w}^{\ell}_j]' \mathbf{V}_j \boldsymbol{\psi}(\bz).
\end{equation}
The preprocessing steps described in this section---fitting a simplified macrotransition entry model \eqref{hazard_preprocess} and performing NMF on the coefficient estimates \eqref{U_factorization}---provide us with basis functions $\phi_{ji}(\bz)$ that we treat as fixed and known during the modeling and inference discussed in Section \ref{sec:Computation}.

Note that an analogous expression for \eqref{all_bases} is used for $\tilde{\xi}^{\ell}_j$ in terms of $\tilde{\mathbf{w}}^{\ell}_j$ and $\tilde{\mathbf{V}}_j$ for pass events; however, for the spatial effect $\xi^{\ell}_\shot$ in the shot probability model, we simply use $\mathbf{V}_5$. Thus, the basis functions for the shot probability model are the same as those for the shot-taking hazard model. 

\subsection{Calculating EPVA: Baseline EPV for League-Average Player}
\label{subsec:EPVA}
To calculate the baseline EPV for a league-average
player possessing the ball in player $\ell$'s shoes, denoted $\nu_t^{r(\ell)}$ in \eqref{EPVA}, we start by considering an alternate version of the transition
probability matrix between coarsened states $\mathbf{P}$. For each
player $\ell_1, \ldots, \ell_5$ on offense, there is a disjoint subset
of rows of $\mathbf{P}$, denoted $\mathbf{P}_{\ell_i}$, that
correspond to possession states for player $\ell_i$. Each row of
$\mathbf{P}_{\ell_i}$ is a probability distribution over transitions
in $\Cset$ given possession in a particular state. Technically, since
states in $\Cset_{\text{poss}}$ encode player identities, players on
different teams do not share all states which they have a nonzero
probability of transitioning to individually. To get around this, we
remove the columns from each $\mathbf{P}_{\ell_i}$ corresponding to
passes to players not on player $\ell_i$'s team, and reorder the
remaining columns according to the position (guard, center, etc.) of
the associated pass recipient. Thus, the interpretation of transition
distributions $\mathbf{P}_{\ell_i}$ across players $\ell_i$ is as
consistent as possible.

We create a baseline transition profile of a hypothetical
league-average player by averaging these transition probabilities across all players: (with
slight abuse of notation) let $\mathbf{P}_r = \sum_{\ell=1}^L
\mathbf{P}_{\ell}/L$. Using this, we create a new transition
probability matrix $\mathbf{P}_r(\ell)$ by replacing player $\ell$'s
transition probabilities ($\mathbf{P}_{\ell}$) with the league-average
player's ($\mathbf{P}_r$). The baseline (league-average) EPV at time
$t$ is then found by evaluating $\nu^{r(\ell)}_t = \E_{\mathbf{P}_r(\ell)}[ X |
C_t]$. Note that $\nu^{r(\ell)}_t$ depends only on the coarsened state $C_t$ at time $t$, rather than the full history of the possession, $\Fz_t$, as in $\nu_t$ \eqref{epveqn}. This ``coarsened'' baseline $\nu_t^{r(\ell)}$ exploits
the fact that, when averaging possessions over the entire season, the
results are (in expectation) identical to using a full-resolution baseline EPV that
assumes the corresponding multiresolution transition probability models
for this hypothetical league-average player. 


